\documentclass[12pt]{article}
\usepackage{enumerate}
\usepackage{amsfonts}
\usepackage{amsmath}
\usepackage{amssymb}

\newcounter{defin}  \newcounter{lemma}  \newcounter{theorem}
\newcounter{property} \newcounter{corol}  \newcounter{remark} \newcounter{example}

\newenvironment{lemma}{\par\refstepcounter{lemma}
     \textbf{Lemma \thelemma.} }{\rm\par}
\newenvironment{theorem}{\par\refstepcounter{theorem}
     \textbf{Theorem \thetheorem.}\ }{\rm\par}
\newenvironment{property}{\par\refstepcounter{property}
     \textbf{Proposition \theproperty.}\ }{\rm\par}
\newenvironment{corollary}{\par\refstepcounter{corol}
     \textbf{Corollary \thecorol.} }{\rm\par}
\newenvironment{definition}{\par\refstepcounter{defin}
     \textbf{Definition \thedefin.}\ }{\rm\par}
\newenvironment{remark}{\par\refstepcounter{remark}
     \textbf{Remark \theremark.}}{\rm\par}
\newenvironment{example}{\par\refstepcounter{example}
     \textbf{Example \theexample.}}{\rm\par}

\newcommand{\s}{|\hspace{-2.7pt}|}

\begin{document}
\title{The output entropy of quantum channels\\ and operations}
\author{M.E.Shirokov\thanks{e-mail:msh@mi.ras.ru}\\Steklov Mathematical Institute, Moscow,
Russia}\date{} \maketitle

\tableofcontents

\pagebreak

\section{Introduction}

The output von Neumann entropy of a quantum channel is an important
characteristic of this channel used in study of its information
properties. This is a concave lower semicontinuous function on the
set of input states of the channel, taking values in $[0,+\infty]$.
In applications, in particular, in analysis of the classical
capacity of a quantum channel it is necessary to have conditions for
continuity of the output entropy of this channel on subsets of input
state space. The natural questions arising in this direction are the
following:
\begin{enumerate}[1)]
    \item What are the conditions under which
the output entropy of a quantum channel is continuous on the whole
space of input states?
    \item What are the conditions under which
the output entropy of a quantum channel is continuous on any set of
input states with continuous entropy?
\end{enumerate}

The first part of this paper is devoted to study of these and some
other questions in the general context of positive linear maps
between Banach spaces of trace-class operators with the special
attention to the classes of quantum channels and operations. In
Section 3 it is shown that finiteness of the output entropy of a
positive linear map on the whole space of input states is equivalent
to its continuity and the sufficient conditions of this property for
a quantum operation and its complementary operation expressed in
terms of Kraus operators are obtained. In Section 4 the
characterization of a positive linear map, for which the property in
the second above-stated question holds, is obtained and its
applications to the class of quantum operations are considered. The
special relation between continuity properties of the output
entropies of a pair of complementary quantum operations and its
corollaries are presented in Section 5. \vspace{5pt}

In the second part of this paper the properties of the output
entropy considered as a function of a pair (map, input state) are
investigated. Such analysis is a necessary tool for exploring
continuity of information characteristics of a quantum channel as
functions of a channel, it also can be used in study of quantum
channels by means of their approximation by quantum operations
\cite{Sh-H,L&S}. In Section 6 the general continuity condition and
the continuity condition based on the complementary relation are
obtained and their corollaries are considered.

In Section 7 the possibilities to prove continuity of the output
entropy of a quantum operation and of a converging sequence of
quantum operations on a \textit{given} set of states, based on the
results of the previous sections, are discussed.

Some applications of the above continuity conditions in analysis of
the Holevo capacity of a quantum channel and of the Entanglement of
Formation of a state of a composite infinite dimensional quantum
system are considered in Section 8.

\section{Preliminaries}

Let $\mathcal{H}$ be a separable Hilbert space,
$\mathfrak{B}(\mathcal{H})$ -- the Banach space of all bounded
operators in $\mathcal{H}$ with the operator norm $\Vert\cdot\Vert$,
$\mathfrak{T}( \mathcal{H})$ -- the Banach space of all trace-class
operators in $\mathcal{H}$ with the trace norm
$\Vert\cdot\Vert_{1}$, containing the cone
$\mathfrak{T}_{+}(\mathcal{H})$ of all positive trace-class
operators. The closed convex subsets
$$
\mathfrak{T}_{1}(\mathcal{H})=\{A\in\mathfrak{T}_{+}(\mathcal{H})\,|\,\mathrm{Tr}A\leq
1\}\;\;\textup{and}\;\;\mathfrak{S}(\mathcal{H})=\{A\in\mathfrak{T}_{+}(\mathcal{H})\,|\,\mathrm{Tr}A=1\}
$$
are complete separable metric spaces with the metric defined by the
trace norm. Operators in $\mathfrak{S}(\mathcal{H})$ are called
density operators or states since each density operator uniquely
defines a normal state on $\mathfrak{B}(\mathcal{H})$.

We will use the Dirac notations $|\varphi\rangle,
|\varphi\rangle\langle\psi|,...$ for vectors $\varphi, \psi,...$ in
a Hilbert space with arbitrary norms (including the zero vector).

In what follows $\mathcal{A}$ is a subset of the cone of positive
trace class operators.

We denote by $\mathrm{cl}(\mathcal{A})$, $\mathrm{co}(\mathcal{A})$,
$\overline{\mathrm{co}}(\mathcal{A})$ and
$\mathrm{extr}(\mathcal{A})$ the closure, the convex hull, the
convex closure and the set of all extreme points of a set
$\mathcal{A}$ correspondingly \cite{J&T,R}.

The set of all Borel probability measures on a closed set
$\mathcal{A}\subseteq\mathfrak{S}(\mathcal{H})$ endowed with the
topology of weak convergence is denoted $\mathcal{P}(\mathcal{A})$.
This set can be considered as a complete separable metric space
\cite{Bogachev,Par}. The \textit{barycenter} $\textbf{b}(\mu)$ of
the measure $\mu$ in $\mathcal{P}(\mathcal{A})$ is the state in
$\overline{\mathrm{co}}(\mathcal{A})$ defined by the Bochner
integral
\[
\textbf{b}(\mu)=\int_{\mathcal{A}}\rho\mu(d\rho).
\]

For arbitrary subset
$\,\mathcal{B}\subseteq\overline{\mathrm{co}}(\mathcal{A})\,$ let
$\mathcal{P}_{\mathcal{B}}(\mathcal{A})$ be the subset of
$\mathcal{P}(\mathcal{A})$ consisting of all measures with
barycenter in $\mathcal{B}$.

Let $\mathcal{P}^{\mathrm{a}}(\mathcal{A})$ be the subset of
$\mathcal{P}(\mathcal{A})$ consisting of atomic measures and let
$\mathcal{P}^{\mathrm{f}}(\mathcal{A})$ be the subset of
$\mathcal{P}^{\mathrm{a}}(\mathcal{A})$ consisting of measures with
a finite number of atoms. Each measure in
$\mathcal{P}^{\mathrm{a}}(\mathcal{A})$ corresponds to a collection
of states $\{\rho_{i}\}\subset\mathcal{A}$ with probability
distribution $\{\pi_{i}\}$ conventionally called \textit{ensemble}
and denoted $\{\pi_{i},\rho _{i}\}$, its barycenter coincides with
the average state $\sum_{i}\pi_{i}\rho_{i}$ of this ensemble.

The identity operator in a Hilbert space $\mathcal{H}$ and the
identity transformation of the Banach space
$\mathfrak{T}(\mathcal{H})$ are denoted respectively
$I_{\mathcal{H}}$ and $\mathrm{Id}_{\mathcal{H}}$.

Let $\mathcal{H}$ and $\mathcal{H}'$ be separable Hilbert spaces
which we call correspondingly input and output space. Let
$\Phi:\mathfrak{T}(\mathcal{H})\rightarrow\mathfrak{T}(\mathcal{H}')$
be a positive linear trace non-increasing map. The \emph{dual} map
$\Phi^{*}:\mathfrak{B}(\mathcal{H}')\rightarrow\mathfrak{B}(\mathcal{H})$
(uniquely defined by the relation
$\mathrm{Tr}\Phi(\rho)A=\mathrm{Tr}\rho\,\Phi^{*}(A),
\rho\in\mathfrak{S}(\mathcal{H}), A\in\mathfrak{B}(\mathcal{H}')$)
is a positive linear map such that $\Phi^{*}(I_{\mathcal{H}'})\leq
I_{\mathcal{H}}$.

The set of positive linear trace non-increasing maps from
$\mathfrak{T}(\mathcal{H})$ to $\mathfrak{T}(\mathcal{H}')$ is
denoted $\mathfrak{L}_{\leq1}^{+}(\mathcal{H},\mathcal{H}')$. The
convex subset of
$\mathfrak{L}_{\leq1}^{+}(\mathcal{H},\mathcal{H}')$, consisting of
completely positive (see \cite{H-SSQT}) maps called \textit{quantum
operations}, is denoted
$\mathfrak{F}_{\leq1}(\mathcal{H},\mathcal{H}')$. The convex subset
of $\mathfrak{F}_{\leq1}(\mathcal{H},\mathcal{H}')$, consisting of
trace preserving  maps called \textit{quantum channels}, is denoted
$\mathfrak{F}_{=1}(\mathcal{H},\mathcal{H}')$.

We assume that the set
$\mathfrak{L}_{\leq1}^{+}(\mathcal{H},\mathcal{H}')$ is endowed with
the topology generated on this set by the strong operator topology
on the set of all linear operators between Banach spaces
$\mathfrak{T}(\mathcal{H})$ and $\mathfrak{T}(\mathcal{H}')$. We
call it \textit{the strong convergence topology}. It is this
topology that makes the sets  $\mathfrak{F}_{\leq
1}(\mathcal{H},\mathcal{H}^{\prime})$ and
$\mathfrak{F}_{=1}(\mathcal{H},\mathcal{H}^{\prime})$ to be
isomorphic to the particular subsets of the cone
$\mathfrak{T}_{+}(\mathcal{H}\otimes\mathcal{H}^{\prime})$ (the
generalized Choi-Jamiolkowski isomorphism) \cite{Sh-H}. Convergence
of a sequence
$\{\Phi_{n}\}\subset\mathfrak{L}_{\leq1}^{+}(\mathcal{H},\mathcal{H}')$
to a map
$\Phi_{0}\in\mathfrak{L}_{\leq1}^{+}(\mathcal{H},\mathcal{H}')$ in
the strong convergence topology means that
$$
\lim_{n\rightarrow+\infty}\Phi_{n}(\rho)=\Phi_{0}(\rho),\quad
\forall\rho\in\mathfrak{S}(\mathcal{H}).
$$

An arbitrary quantum operation (correspondingly channel)
$\Phi\in\mathfrak{F}_{\leq 1}(\mathcal{H},\mathcal{H}^{\prime})$ has
the following Kraus representation
\begin{equation}\label{Kraus-rep}
\Phi(\cdot)=\sum_{i=1}^{+\infty}V_{i}(\cdot)V^{*}_{i},
\end{equation}
where $\{V_{i}\}_{i=1}^{+\infty}$ is a set of bounded linear
operators from $\mathcal{H}$ into $\mathcal{H}'$ such that
$\sum_{i=1}^{+\infty}V^{*}_{i}V_{i}\leq I_{\mathcal{H}}$
(correspondingly
$\sum_{i=1}^{+\infty}V^{*}_{i}V_{i}=I_{\mathcal{H}}$).\vspace{5pt}

For natural $k$ we denote by $\mathfrak{F}^{k}_{\leq
1}(\mathcal{H},\mathcal{H}^{\prime})$ (correspondingly by
$\mathfrak{F}^{k}_{=1}(\mathcal{H},\mathcal{H}^{\prime})$) the
subset of  $\mathfrak{F}_{\leq 1}(\mathcal{H},\mathcal{H}^{\prime})$
consisting of quantum operations (correspondingly of quantum
channels) having the Kraus representation with $\leq k$ nonzero
summands.\vspace{5pt}

If $\Phi$ is a quantum operation (correspondingly channel) in
$\mathfrak{F}_{\leq1}(\mathcal{H},\mathcal{H}^{\prime})$ then by the
Stinespring dilation theorem there exist a Hilbert space
$\mathcal{H}''$ and a contraction (correspondingly isometry)
$V:\mathcal{H}\rightarrow\mathcal{H}'\otimes\mathcal{H}''$ such that
\begin{equation}\label{Stinespring-rep}
\Phi(A)=\mathrm{Tr}_{\mathcal{H}''}VA V^{*},\quad \forall
A\in\mathfrak{T}(\mathcal{H}).
\end{equation}
The quantum operation (correspondingly channel)
\begin{equation}\label{c-channel}
\mathfrak{T}(\mathcal{H})\ni
A\mapsto\widetilde{\Phi}(A)=\mathrm{Tr}_{\mathcal{H}'}VAV^{*}\in\mathfrak{T}(\mathcal{H}'')
\end{equation}
is called \emph{complementary} to the operation (correspondingly
channel) $\Phi$ \cite{D&Sh}.\footnote{The operation
$\widetilde{\Phi}$ is also called \emph{conjugate} or
\emph{canonically dual} to the operation $\Phi$ \cite{KMNR,Winter}.}
\vspace{5pt}

We will use the following extension of the von Neumann entropy
\break $H(\rho)=-\mathrm{Tr}\rho\log\rho$ of a state
$\rho\in\mathfrak{S}(\mathcal{H})$ to the cone
$\mathfrak{T}_{+}(\mathcal{H})$ (cf.\cite{L-2})
$$
H(A)=\mathrm{Tr}\eta(A)-\eta(\mathrm{Tr}A),\quad \forall A
\in\mathfrak{T}_{+}(\mathcal{H}),\quad \textrm{where}\;\;
\eta(x)=-x\log x.
$$

In what follows the function $A\mapsto H(A)$ on the cone
$\mathfrak{T}_{+}(\mathcal{H})$ is called the \textit{quantum
entropy} while the function $\{x_i\}\mapsto
H(\{x_i\})=\sum_{i}\eta(x_{i})-\eta\left(\sum_{i}x_{i}\right)$ on
the positive cone of the space $\ell_{1}$, coinciding with the
Shannon entropy on the set $\mathfrak{P}_{+\infty}$ of probability
distributions, is called the \textit{classical entropy}.\vspace{5pt}

Nonnegativity, concavity and lower semicontinuity of the von Neumann
entropy on the set $\mathfrak{S}(\mathcal{H})$ imply the same
properties of the quantum entropy on the set
$\mathfrak{T}_{+}(\mathcal{H})$. By definition we have
\begin{equation}\label{H-fun-eq}
H(\lambda A)=\lambda H(A),\quad
A\in\mathfrak{T}_{+}(\mathcal{H}),\;\lambda\geq 0.
\end{equation}
This relation and  proposition 6.2 in \cite{O&P} imply
\begin{equation}
H(A)+H(B-A)\leq H(B)\leq H(A)+H(B-A)+\mathrm{Tr}B
h_{2}\left(\frac{\mathrm{Tr}
A}{\mathrm{Tr}B}\right)\label{H-fun-ineq},
\end{equation}
where $A,B\in\mathfrak{T}_{+}(\mathcal{H}),\; A\leq B,$ and
$\,h_{2}(x)=\eta(x)+\eta(1-x)$.\vspace{5pt}

By using theorem 11.10 in \cite{N&Ch} and a simple approximation it
is easy to obtain the following inequality
\begin{equation}\label{w-k-ineq}
\sum_{i=1}^{n}\lambda_{i}H(A_{i})\leq
H\left(\sum_{i=1}^{n}\lambda_{i}A_{i}\right)\leq
\sum_{i=1}^{n}\lambda_{i}H(A_{i})+H\left(\{\lambda_{i}\}_{i=1}^{n}\right),
\end{equation}
valid for any set $\{A_{i}\}_{i=1}^{n}$ of operators in
$\mathfrak{T}_{1}(\mathcal{H})$ and any probability distribution
$\{\lambda_{i}\}_{i=1}^{n}$, where $n\leq+\infty$. This inequality
implies the following one
\begin{equation}\label{w-k-ineq+}
\sum_{i=1}^{n}H(A_{i})\leq H\left(\sum_{i=1}^{n}A_{i}\right)\leq
\sum_{i=1}^{n}H(A_{i})+H\left(\left\{\mathrm{Tr}A_{i}\right\}_{i=1}^{n}\right),
\end{equation}
valid for any set $\{A_{i}\}_{i=1}^{n}$ of operators in
$\mathfrak{T}_{1}(\mathcal{H})$ with finite
$\sum_{i=1}^{n}\!\mathrm{Tr}A_{i}$. \vspace{5pt}

Following \cite{H-Sh-2} an arbitrary positive unbounded operator in
a separable Hilbert space with discrete spectrum of finite
multiplicity is called $\mathfrak{H}$-\textit{operator}. Let
$\mathrm{g}(H)=\inf\{\lambda>0\,|\,\mathrm{Tr}e^{-\lambda
H}<+\infty\}$ assuming that $\mathrm{g}(H)=+\infty$ if
$\mathrm{Tr}e^{-\lambda H}=+\infty$ for all $\lambda>0$. For given
$\mathfrak{H}$-operator $H$ in a Hilbert space $\mathcal{H}$ and
positive $h$ consider the convex set
$$
\mathcal{K}_{H,h}=\{A\in\mathfrak{T}_{1}(\mathcal{H})\,|\,\mathrm{Tr}AH\leq
h\}.
$$

We will use the following generalizations\footnote{These
generalizations can be easily obtained by using the construction
from the proof of Lemma \ref{H-oper} below.} of proposition 1a in
\cite{Sh-4} and proposition 6.6. in \cite{O&P}.\vspace{5pt}

\begin{property}\label{H-cont-cond} \emph{Let $H$ be a $\mathfrak{H}$-operator in a Hilbert space $\mathcal{H}$ and
$h>0$.}
\begin{enumerate}[A)\!\!]
    \item \textit{The quantum entropy
is bounded on the set $\mathcal{K}_{H,h}$ if and only if
$\mathrm{g}(H)\!<\!+\infty$};
    \item \textit{The quantum entropy
is continuous on the set $\mathcal{K}_{H,h}$ if and only if
$\mathrm{g}(H)\!=\!0$}.
\end{enumerate}
\end{property}\vspace{5pt}

We will also use the following result easily derived from
corollaries 3 and
 4 in \cite{Sh-11}.\vspace{5pt}

\begin{lemma}\label{face}
\emph{Let $\{A_{n}\}$ and $\{B_{n}\}$ be  sequences of operators in
$\mathfrak{T}_{+}(\mathcal{H})$ converging respectively to operators
$A_{0}$ and $B_{0}$. Then
$$
\lim_{n\rightarrow+\infty}H(A_{n}+B_{n})=H(A_{0}+B_{0})
$$
if and only if
$\quad\displaystyle\lim_{n\rightarrow+\infty}H(A_{n})=H(A_{0})\quad
and \quad \lim_{n\rightarrow+\infty}H(B_{n})=H(B_{0})$.}
\end{lemma}
\vspace{5pt}

The quantum entropy of an arbitrary operator
$A\in\mathfrak{T}_{+}(\mathcal{H})$ and the classical entropy of the
sequence of its diagonal values in any orthonormal basis
$\{|i\rangle\}_{i=1}^{+\infty}$ of the space $\mathcal{H}$ are
related as follows
\begin{equation}\label{H-q-c}
H(A)\leq H\left(\left\{\langle
i|A|i\rangle\right\}_{i=1}^{+\infty}\right)
\end{equation}
(this inequality can be proved by using nonnegativity of the
relative entropy).\vspace{5pt}

By using relations (\ref{H-fun-eq}) and (\ref{H-q-c}) it is easy to
derive from proposition 5E in \cite{Sh-4} the following continuity
condition for the quantum entropy.\vspace{5pt}

\begin{property}\label{cont-cond-a-2}
\textit{Let $\{|i\rangle\}_{i=1}^{+\infty}$ be an orthonormal basis
of a Hilbert space $\mathcal{H}$. Continuity of the quantum entropy
on a set $\mathcal{A}\subset\mathfrak{T}_{+}(\mathcal{H})$ follows
from continuity of the classical entropy on the set
$\,\left\{\{\langle
i|A|i\rangle\}_{i=1}^{+\infty}\,|\,A\in\mathcal{A}\right\}\subset
(\ell_{1})_{+}$.}
\end{property}\vspace{5pt}

We will use the triangle inequality
\begin{equation}\label{d-ineq}
H(C)\geq\left|H\left(\mathrm{Tr}_{\mathcal{K}}C\right)-H\left(\mathrm{Tr}_{\mathcal{H}}C\right)\right|
\end{equation}
valid for any operator $C$ in
$\mathfrak{T}_{+}(\mathcal{H}\otimes\mathcal{K})$
\cite{N&Ch}.\vspace{5pt}

For arbitrary map $\Phi\in\mathfrak{L}^{+}_{\leq
1}(\mathcal{H},\mathcal{H}')$ and  operator
$A\in\mathfrak{T}_{+}(\mathcal{H})$ the following estimation holds
\begin{equation}\label{out-ent-est}
H(\Phi(A))\leq
\left[\sup_{\rho\in\mathrm{extr}\mathfrak{S}(\mathcal{H})}H(\Phi(\rho))\right]\mathrm{Tr}A+H(A),
\end{equation}
which is proved by using the spectral decomposition of $A$ and
inequality (\ref{w-k-ineq}).\vspace{5pt}

The relative entropy for two operators $A$ and $B$ in
$\mathfrak{T}_{+}(\mathcal{H})$ is defined as follows
(cf.\cite{L-2})
$$
H(A\,\|B)=\sum_{i=1}^{+\infty}\langle i|\,(A\log A-A\log
B+B-A)\,|i\rangle,
$$
where $\{|i\rangle\}_{i=1}^{+\infty}$ is the orthonormal basis of
eigenvectors of the operator $A$ and it is assumed that
$H(A\,\|B)=+\infty$ if $\,\mathrm{supp}A$ is not contained in
$\mathrm{supp}B$.\vspace{5pt}

For natural $k$ we denote by $\mathfrak{T}_{+}^{k}(\mathcal{H})$
(correspondingly by $\mathfrak{S}_{k}(\mathcal{H})$) the set of
positive trace class operators (correspondingly states) having rank
$\leq k$.\vspace{5pt}

For natural $k$ and lower bounded Borel function $f$ on the set
$\mathfrak{S}_{k}(\mathcal{H})$ consider the functions
\begin{equation}\label{sigma-roof}
\mathfrak{S}(\mathcal{H})\ni
\rho\mapsto\hat{f}_{k}^{\sigma}(\rho)=\sup_{\{\pi_{i},\rho_{i}\}\in\mathcal{P}^{\mathrm{a}}_{\{\rho\}}(\mathfrak{S}_{k}(\mathcal{H}))}
\sum_{i}\pi_{i}f(\rho_{i})
\end{equation}
and
\begin{equation}\label{mu-roof}
\mathfrak{S}(\mathcal{H})\ni
\rho\mapsto\hat{f}_{k}^{\mu}(\rho)=\sup_{\mu\in\mathcal{P}_{\{\rho\}}(\mathfrak{S}_{k}(\mathcal{H}))}
\int_{\mathfrak{S}_{k}(\mathcal{H})}f(\sigma)\mu(d\sigma)
\end{equation}
(the first supremum is over all decompositions of the state $\rho$
into countable convex combination of states of rank $\leq k$, the
second one is over all probability measures with the barycenter
$\rho$ supported by states of rank $\leq k$). Properties of these
functions are studied in \cite{Sh-11}, where it is shown that lower
semicontinuity of the function $f$ on the set
$\mathfrak{S}_{k}(\mathcal{H})$ implies lower semicontinuity and
coincidence of the functions $\hat{f}_{k}^{\sigma}$ and
$\hat{f}_{k}^{\mu}$ on the set
$\mathfrak{S}(\mathcal{H})$.\vspace{5pt}

Following \cite{Sh-11} we say that a subset $\mathcal{A}$ of the
cone $\mathfrak{T}_{+}(\mathcal{H})$ of positive trace class
operators has the uniform approximation property (the
UA\nobreakdash-\hspace{0pt}property) if
$$
\lim_{k\rightarrow+\infty}\sup_{A\in\mathcal{A}}\Delta_{k}(A)=0,
$$
where
$$
\Delta_{k}(A)=\inf_{\{\pi_{i},A_{i}\}\in\mathcal{P}^{\mathrm{a}}_{\{A\}}(\mathfrak{T}_{+}^{k}(\mathcal{H}))}
\sum_{i}\pi_{i}H(A_{i}\|A)
$$
for each natural $k$ (the infimum is over all decompositions of the
operator $A$ into countable convex combination of operators of rank
$\leq k$). The UA\nobreakdash-\hspace{0pt}property of a set
$\mathcal{A}$ is a sufficient condition for continuity of the
quantum entropy on this set, if the set $\mathcal{A}$ is compact
then this condition is also necessary. \vspace{5pt}

We will use the following simple observation
\begin{equation}\label{u-h-c}
\lim_{y\rightarrow0}\sup_{x\in[0,1]}(x+y)h_{2}\left(\frac{y}{x+y}\right)=0.
\end{equation}

\textbf{Note:}  In what follows  continuity of a function $f$ on a
set $\mathcal{A}\subset\mathfrak{T}_{+}(\mathcal{H})$ implies its
finiteness on this set (in contrast to lower (upper)
semicontinuity).

\section{On continuity of the output entropy of\\ a positive linear map}

\subsection{The general case}

Let
$\,\Phi:\mathfrak{T}(\mathcal{H})\rightarrow\mathfrak{T}(\mathcal{H}')\,$
be a positive linear map. The output entropy $H_{\Phi}\doteq
H\circ\Phi$ of this map is a concave nonnegative lower
semicontinuous function on the set
$\mathfrak{S}(\mathcal{H})\subset\mathfrak{T}(\mathcal{H})$. The
following theorem shows that this function can not be finite and
discontinuous simultaneously. \vspace{50pt}

\begin{theorem}\label{bound-cont}
\textit{Let $\,\Phi$ be a map in
$\mathfrak{L}_{\leq1}^{+}(\mathcal{H},\mathcal{H}')$. The following
properties are equivalent:}
\begin{enumerate}[(i)]
    \item \textit{the function $\rho\mapsto H_{\Phi}(\rho)$ is finite on  $\,\mathfrak{S}(\mathcal{H})$;}
    \item \textit{the function $\rho\mapsto H_{\Phi}(\rho)$ is continuous and bounded on  $\,\mathfrak{S}(\mathcal{H})$;}
    \item \textit{there exists an orthonormal basis $\{|i\rangle\}_{i=1}^{+\infty}$ of the space $\mathcal{H}'$ such that
  the function $\rho\mapsto H\left(\left\{\langle i|\Phi(\rho)|i\rangle\right\}_{i=1}^{+\infty}\right)$
  is continuous and bounded on
  $\,\mathfrak{S}(\mathcal{H})$;}\footnote{By Proposition \ref{cont-cond-a-2} this property is formally stronger than the previous one.}

  \item \textit{there exist an orthonormal basis $\{|i\rangle\}_{i=1}^{+\infty}$ of the space $\mathcal{H}'$ and a sequence $\{h_{i}\}_{i=1}^{+\infty}$ of
  nonnegative numbers such that
$$
\|\sum_{i=1}^{+\infty}h_{i}\Phi^{*}(|i\rangle\langle
i|)\|<+\infty\quad\textit{and}\quad
\sum_{i=1}^{+\infty}e^{-h_{i}}<+\infty,
$$
where
$\Phi^{*}:\mathfrak{B}(\mathcal{H}')\rightarrow\mathfrak{B}(\mathcal{H})$
is the dual map to the map $\Phi$.}
\end{enumerate}
\emph{The set $\,\mathfrak{S}(\mathcal{H})$ in $\,\mathrm{(i)}$ can
be replaced by arbitrary convex closed bounded subset
$\mathcal{A}\subset\mathfrak{T}_{+}(\mathcal{H})$ such that
$\;\sup_{A\in\mathcal{A}}\lim_{n\rightarrow+\infty}\mathrm{Tr}AB_{n}<+\infty\;\Rightarrow\;\sup_{n}\|B_{n}\|<+\infty$
for any increasing sequence $\{B_{n}\}$ of positive operators in
$\,\Phi^{*}(\mathfrak{B}(\mathcal{H}'))$.}

\end{theorem} \vspace{5pt}

\textbf{Proof.} $\mathrm{(i)\Rightarrow(ii)}$ Validity of the
discrete Jensen inequality for the concave finite nonnegative
function $\rho\mapsto H_{\Phi}(\rho)$ implies its boundedness.
Indeed, if for each natural $n$ there exists a state $\rho_{n}$ such
that $H_{\Phi}(\rho_{n})\geq 2^{n}$ then
$$
H_{\Phi}\left(\sum_{n=1}^{+\infty}2^{-n}\rho_{n}\right)\geq\sum_{n=1}^{+\infty}2^{-n}H_{\Phi}\left(\rho_{n}\right)=+\infty.
$$

Lemma \ref{H-oper} below implies existence of a
$\mathfrak{H}$\nobreakdash-\hspace{0pt}operator $H=-\log T$ such
that $\mathrm{g}(H)<+\infty$ and $\mathrm{Tr}H\Phi(\rho)\leq h$ for
all $\rho\in\mathfrak{S}(\mathcal{H})$ and some $h>0$. Let
$H=\sum_{i=1}^{+\infty}h_{i}|i\rangle\langle i|$. Since the function
$$
\rho\mapsto\mathrm{Tr}H\Phi(\rho)=\sum_{i=1}^{+\infty}h_{i}\langle
i|\Phi(\rho)|i\rangle=\mathrm{Tr}\left[\sum_{i=1}^{+\infty}h_{i}\Phi^{*}(|i\rangle\langle
i|)\right]\rho
$$
is bounded on $\mathfrak{S}(\mathcal{H})$, the linear operator in
the squire brackets is bounded on $\mathcal{H}$. Thus the above
function is continuous on $\mathfrak{S}(\mathcal{H})$. For arbitrary
compact subset $\mathcal{C}$ of $\mathfrak{S}(\mathcal{H})$ Dini's
lemma implies uniform convergence of the series
$\sum_{i=1}^{+\infty}h_{i}\mathrm{Tr}\Phi^{*}(|i\rangle\langle
i|)\rho$ on the set $\mathcal{C}$ and hence existence of a
nondecreasing sequence $\{y^{\mathcal{C}}_{i}\}_{i=1}^{+\infty}$ of
positive numbers converging to $+\infty$ such that
$\sup_{\rho\in\mathcal{C}}\sum_{i=1}^{+\infty}y^{\mathcal{C}}_{i}h_{i}\mathrm{Tr}\Phi^{*}(|i\rangle\langle
i|)\rho<+\infty$. Let
$H^{\mathcal{C}}=\sum_{i=1}^{+\infty}y^{\mathcal{C}}_{i}h_{i}|i\rangle\langle
i|$ be a $\mathfrak{H}$\nobreakdash-\hspace{0pt}operator with
$\mathrm{g}(H^{\mathcal{C}})=0$. Thus we have
\begin{equation}\label{int-exp}
\sup_{\rho\in\mathcal{C}}\mathrm{Tr}H^{\mathcal{C}}\Phi(\rho)
=\sup_{\rho\in\mathcal{C}}\sum_{i=1}^{+\infty}y^{\mathcal{C}}_{i}h_{i}\mathrm{Tr}\Phi^{*}(|i\rangle\langle
i|)\rho<+\infty.
\end{equation}
By proposition \ref{H-cont-cond}B the function $\rho\mapsto
H(\Phi(\rho))$ is continuous on the set $\mathcal{C}$ and hence on
the set $\mathfrak{S}(\mathcal{H})$ (since $\mathcal{C}$ is an
arbitrary compact subset of $\mathfrak{S}(\mathcal{H})$).

$\mathrm{(i)\Rightarrow(iv)}$ In the proof of
$\mathrm{(i)\Rightarrow(ii)}$ existence of the basis
$\{|i\rangle\}_{i=1}^{+\infty}$ and of the sequence
$\{h'_{i}=\lambda h_{i}\}_{i=1}^{+\infty},\,\lambda>0,$ with the
desired properties is shown.

$\mathrm{(iv)\Rightarrow(iii)}$ follows from the proof of
$\mathrm{(i)\Rightarrow(ii)}$ since (\ref{int-exp}) and Proposition
\ref{H-cont-cond}B implies continuity of the function $\rho\mapsto
H\left(\left\{\langle
i|\Phi(\rho)|i\rangle\right\}_{i=1}^{+\infty}\right)$ on the set
$\mathcal{C}$.

$\mathrm{(iii)\Rightarrow(i)}$ follows from relation (\ref{H-q-c}).

The last assertion of the theorem is a corollary of the proof of
$\mathrm{(i)\Rightarrow(ii)}$. $\square$ \vspace{5pt}

\begin{lemma}\label{H-oper}
\emph{For an arbitrary convex set
$\mathcal{A}\subset\mathfrak{T}_{1}(\mathcal{H})$, on which the
quantum entropy is bounded, there exists an operator
$\;T\in\mathfrak{T}_{1}(\mathcal{H})$ such that
$$
\sup_{A\in\mathcal{A}}\mathrm{Tr}A(-\log T)<+\infty\quad
\textit{and} \quad UT=TU
$$
for any unitary $\;U$ in $\mathfrak{B}(\mathcal{H})$ such that
$\;UAU^{*}\in\mathcal{A}$ for all $A\in\mathcal{A}$.}
\end{lemma}\vspace{5pt}

\textbf{Proof.} Let $\mathcal{K}$ be the one dimensional space
generated by the unit vector $|0\rangle$. Consider the convex set
$$
\mathcal{A}^{\mathrm{e}}=\left\{\rho_{A}=A\oplus(1-\mathrm{Tr}A)|0\rangle\langle
0|\,|\,A\in\mathcal{A}\right\}
$$
of states in $\mathfrak{S}(\mathcal{H}\oplus\mathcal{K})$. For
arbitrary $A\in\mathcal{A}$ we have
$$
H(\rho_{A})=-\mathrm{Tr}A\log A+\eta(1-\mathrm{Tr}A)
=H(A)+\eta(\mathrm{Tr}A)+\eta(1-\mathrm{Tr}A)\leq H(A)+1.
$$
Thus the von Neumann entropy is bounded on the convex set
$\mathcal{A}^{\mathrm{e}}$. Hence the
$\chi$\nobreakdash-\hspace{0pt}capacity
$\bar{C}(\mathcal{A}^{\mathrm{e}})$ of this set is finite
\cite{Sh-4}. Theorem 1 in \cite{Sh-4} implies existence of the
unique state $\Omega(\mathcal{A}^{\mathrm{e}})$ in
$\mathrm{cl}(\mathcal{A}^{\mathrm{e}})$ (called the optimal average
state of the set $\mathcal{A}^{\mathrm{e}}$) such that
$$
H(\rho\|\Omega(\mathcal{A}^{\mathrm{e}}))\leq
\bar{C}(\mathcal{A}^{\mathrm{e}})
$$
for all $\rho\in\mathcal{A}^{\mathrm{e}}$. The state
$\Omega(\mathcal{A}^{\mathrm{e}})$ has the form $T\oplus
\lambda|0\rangle\langle 0|$, where
$T\in\mathfrak{T}_{1}(\mathcal{H})$ and $\lambda>0$.

For arbitrary unitary $U$ in $\mathfrak{B}(\mathcal{H})$ such that
$U\mathcal{A}U^{*}=\mathcal{A}$ corollary 8 in \cite{Sh-4} implies
$(U\oplus
I_{\mathcal{K}})\Omega(\mathcal{A}^{\mathrm{e}})=\Omega(\mathcal{A}^{\mathrm{e}})(U\oplus
I_{\mathcal{K}})$ and hence $UT=TU$.  $\square$ \vspace{5pt}

\begin{remark}\label{on-cont} Theorem \ref{bound-cont} \emph{do not assert}
that finiteness of the quantum entropy on the set
$\Phi(\mathfrak{S}(\mathcal{H}))$ implies its continuity on this set
since continuity of the function $\rho\mapsto H_{\Phi}(\rho)\doteq
H(\Phi(\rho))$ on the \emph{noncompact} set
$\mathfrak{S}(\mathcal{H})$ does not imply continuity of the
function $A\mapsto H(A)$ on the set
$\Phi(\mathfrak{S}(\mathcal{H}))$. To show this consider the
following example.

Let $\mathcal{A}$ be a convex closed subset of
$\mathfrak{S}(\mathcal{H}')$ on which the von Neumann entropy is
bounded but not continuous (see the examples in \cite{Sh-4}). Let
$\{\sigma_{n}\}_{n=1}^{+\infty}$ be a sequence of states in
$\mathcal{A}$ converging to a state $\sigma_{0}$  such that
$\lim_{n\rightarrow+\infty}H(\sigma_{n})\neq H(\sigma_{0})$.
Consider the map $\Phi:\rho\mapsto\sum_{n\geq0}\langle
n|\rho|n\rangle\sigma_{n}$, where $\{|n\rangle\}_{n\geq0}$ is a
particular orthonormal basis in $\mathcal{H}$. By Theorem
\ref{bound-cont} the function $\rho\mapsto H_{\Phi}(\rho)$ is
continuous on the set $\mathfrak{S}(\mathcal{H})$ but the function
$A\mapsto H(A)$ is not continuous on the set
$\Phi(\mathfrak{S}(\mathcal{H}))$ containing the sequence
$\{\sigma_{n}\}_{n=1}^{+\infty}$ and the state $\sigma_{0}$.

Continuity of the function $\rho\mapsto H_{\Phi}(\rho)$ on the set
$\mathfrak{S}(\mathcal{H})$  means continuity of the function
$A\mapsto H(A)$ on each set of the form $\Phi(\mathcal{C})$, where
$\mathcal{C}$ is a compact subset of $\mathfrak{S}(\mathcal{H})$.
\end{remark}\vspace{5pt}

\begin{remark}\label{on-cont+}
The main assertion of Theorem \ref{bound-cont} (the implication
$\mathrm{(i)\Rightarrow(ii)}$) is based on the specific property of
the von Neumann entropy, it can not be proved by using only such
general properties of  entropy-type functions as concavity, lower
semicontinuity and nonnegativity. The simplest example showing this
is given by the function $\rho\mapsto R_{0}(\Phi(\rho))\doteq
\|\Phi(\rho)\|_{1}\log\mathrm{rank}(\Phi(\rho))$ -- the output
$0$\nobreakdash-\hspace{0pt}order Renyi entropy of the map $\Phi$.
\end{remark}\vspace{5pt}

Theorem \ref{bound-cont} can be used to obtain a condition of
continuity of the output entropy for the following class of quantum
channels.\vspace{5pt}

\begin{example}\label{e-b-c}
Let $G$ be a compact group, $\{V_{g}\}_{g\in G}$ be an unitary
representation of $G$ on $\mathcal{H}'$, $M$ be a positive
operator-valued measure (POVM) on $G$ taking values in
$\mathfrak{B}(\mathcal{H})$.  For given arbitrary state $\sigma$ in
$\mathfrak{S}(\mathcal{H}')$ consider the quantum channel
\begin{equation*}
\Phi_{\sigma}(\rho)=\int_{G}V_{g}\sigma V_{g}^{*}\mathrm{Tr}\rho
M(dg).
\end{equation*}%
The channel of this type was used in \cite{H-Sh-W} as an example of
entanglement-breaking channel which has no Kraus representation with
operators of rank~1.\footnote{An arbitrary finite dimensional
entanglement-breaking channel has Kraus  representation with
operators of rank~1~\cite{Ruskai}.}

By Theorem \ref{bound-cont} the channel $\Phi_{\sigma}$ has
continuous output entropy if the state
$\omega(G,V_{g},\sigma)=\int_{G}V_{g}\sigma V_{g}^{*}\mu_{H}(dg)$,
where $\mu_{H}$ is the Haar measure on $G$, has finite entropy. This
condition is necessary if the set of probability measures
$\{\mathrm{Tr}\rho M(\cdot)\}_{\rho\in\mathfrak{S}(\mathcal{H})}$ is
weakly dense in the set of all probability measures on $G$.
\end{example}\vspace{5pt}

Theorem \ref{bound-cont} and inequality (\ref{w-k-ineq}) imply the
following observation (which can be directly proved by using Lemma
\ref{face}).\vspace{5pt}
\begin{corollary}\label{bound-cont-c}
\textit{Let $\,\Phi$ and $\,\Psi$ be  maps in
$\,\mathfrak{L}_{\leq1}^{+}(\mathcal{H},\mathcal{H}')$ and
$\lambda\in(0,1)$. The map $\,\lambda\Phi+(1-\lambda)\Psi$ has
continuous output entropy if and only if the maps $\,\Phi$ and
$\,\Psi$ have continuous output entropy.}
\end{corollary}\vspace{5pt}

Thus the set of all positive maps with continuous output entropy is
convex and forms a \emph{face} of the convex set
$\,\mathfrak{L}_{\leq1}^{+}(\mathcal{H},\mathcal{H}')$. It is easy
to show that this face is dense in
$\,\mathfrak{L}_{\leq1}^{+}(\mathcal{H},\mathcal{H}')$ (in the
strong convergence topology).\vspace{5pt}

We will use the following corollary of Theorem \ref{bound-cont} and
inequality (\ref{w-k-ineq+}).\vspace{5pt}

\begin{corollary}\label{bound-cont-c+}
\emph{Let $\,\{\Phi_{i}\}_{i\in I}$ be a finite or countable family
of maps in \break
$\mathfrak{L}_{\leq1}^{+}(\mathcal{H},\mathcal{H}')$ such that
$\,\sup_{\rho\in\mathfrak{S}(\mathcal{H})}\sum_{i\in
I}\mathrm{Tr}\Phi_{i}(\rho)<+\infty$. The output entropy of the map
$\sum_{i\in I}\Phi_{i}$ is continuous if
$$
\sum_{i\in I}H(\Phi_{i}(\rho))<+\infty\quad\textit{and}\quad
H\left(\left\{\mathrm{Tr}\Phi_{i}(\rho)\right\}_{i\in
I}\right)<+\infty\quad \forall\rho\in\mathfrak{S}(\mathcal{H}).
$$
This condition is necessary if either
$\,\mathrm{supp}\,\Phi_{i}(\rho)\perp\mathrm{supp}\,\Phi_{j}(\rho)$
for all $i\neq j$ and all $\rho\in\mathfrak{S}(\mathcal{H})$ or the
set $I$ is finite.}
\end{corollary}\vspace{5pt}

Theorem \ref{bound-cont} provides a simple proof of the following
result.\footnote{It can be also proved by using Proposition
\ref{cont-cond-a-1} in the Appendix and corollary 4 in
\cite{Sh-11}.}\vspace{5pt}

\begin{corollary}\label{tensor-pr} \textit{Let
$\,\Phi:\mathfrak{T}(\mathcal{H})\rightarrow\mathfrak{T}(\mathcal{H}')$
and
$\,\Psi:\mathfrak{T}(\mathcal{K})\rightarrow\mathfrak{T}(\mathcal{K}')$
be two positive linear bounded maps having continuous output
entropy. If the map
$\,\Phi\otimes\Psi:\mathfrak{T}(\mathcal{H}\otimes\mathcal{K})\rightarrow\mathfrak{T}(\mathcal{H}'\otimes\mathcal{K}')$
is positive then it has continuous output entropy.}
\end{corollary}\vspace{5pt}

\textbf{Proof.} We may assume that the maps $\Phi$ and $\Psi$ are
trace non-increasing. By Theorem \ref{bound-cont} it is sufficient
to prove that $H(\Phi\otimes\Psi(\omega))<+\infty$ for any
$\omega\in\mathfrak{S}(\mathcal{H}\otimes\mathcal{K})$. This follows
from subadditivity of the quantum entropy since
$\mathrm{Tr}_{\mathcal{K}}\Phi\otimes\Psi(\omega)\leq\Phi(\mathrm{Tr}_{\mathcal{K}}\omega)$
and
$\mathrm{Tr}_{\mathcal{H}}\Phi\otimes\Psi(\omega)\leq\Psi(\mathrm{Tr}_{\mathcal{H}}\omega)$.
$\square$\vspace{5pt}

In Section 4 we will use the following corollary of Theorem
\ref{bound-cont}.\vspace{5pt}
\begin{corollary}\label{tensor-pr+} \textit{Let
$\,\Phi:\mathfrak{T}(\mathcal{H})\rightarrow\mathfrak{T}(\mathcal{H}')$
and
$\,\Psi:\mathfrak{T}(\mathcal{K})\rightarrow\mathfrak{T}(\mathcal{K}')$
be positive linear bounded maps such that the map
$\,\Phi\otimes\Psi:\mathfrak{T}(\mathcal{H}\otimes\mathcal{K})\rightarrow\mathfrak{T}(\mathcal{H}'\otimes\mathcal{K}')$
is positive ($\mathcal{H},\mathcal{H}',\mathcal{K}$, $\mathcal{K}'$
are separable Hilbert spaces). If the map $\,\Psi$ is trace
preserving and has finite (and hence continuous) output entropy then
the following properties are equivalent:}
\begin{enumerate}[(i)]
    \item
   \emph{ $H(\Phi\otimes\Psi(|\varphi\rangle\langle\varphi|))<+\infty$ for
    any unit vector
    $\varphi\in\mathcal{H}\otimes\mathcal{K}$;}
    \item \emph{the map $\Phi$ has
continuous output entropy;}
   \item \emph{the map $\Phi\otimes\Psi$ has
continuous output entropy.}
\end{enumerate}
\emph{If the map $\,\Phi$ is trace preserving then the condition of
finiteness of the output entropy of the map $\,\Psi$  can be
replaced by the condition}
$$
\min\left\{H_{\Phi}(\mathrm{Tr}_{\mathcal{K}}|\varphi\rangle\langle\varphi|),H_{\Psi}(\mathrm{Tr}_{\mathcal{H}}|\varphi\rangle\langle\varphi|)\right\}
<+\infty\quad\forall\varphi\in\mathcal{H}\otimes\mathcal{K}.
$$
\end{corollary}\vspace{5pt}

\textbf{Proof.} We may assume that the map $\Phi$ is trace
non-increasing.

$\mathrm{(i)\Rightarrow(ii)}$ Let $\rho$ be an arbitrary state in
$\mathfrak{S}(\mathcal{H})$ and $|\varphi\rangle$ be a vector in
$\mathcal{H}\otimes\mathcal{K}$ such that
$\rho=\mathrm{Tr}_{\mathcal{K}}|\varphi\rangle\langle\varphi|$.
Since the map $\Psi$ is trace preserving we have
$\mathrm{Tr}_{\mathcal{K}}\Phi\otimes\Psi(|\varphi\rangle\langle\varphi|)=\Phi(\rho)$.
By noting that
$\mathrm{Tr}_{\mathcal{H}}\Phi\otimes\Psi(|\varphi\rangle\langle\varphi|)\leq\Psi(\sigma)$,
where
$\sigma=\mathrm{Tr}_{\mathcal{H}}|\varphi\rangle\langle\varphi|$,
and by using finiteness of $H(\Psi(\sigma))$ with (\ref{H-fun-ineq})
and (\ref{d-ineq}) we conclude that $H(\Phi(\rho))<+\infty$. By
Theorem \ref{bound-cont} the map $\Phi$ has continuous output
entropy.

$\mathrm{(ii)\Rightarrow(iii)}$ follows from Corollary
\ref{tensor-pr}.

$\mathrm{(iii)\Rightarrow(i)}$ is obvious.

The last assertion of the corollary is proved by the similar
argumentation. $\square$ \vspace{5pt}

Property $\mathrm{(iv)}$ in Theorem \ref{bound-cont} can be
considered as a criterion of continuity of the output entropy of a
map $\Phi$ in terms of its dual map $\Phi^{*}$. It will be used in
the proof of Proposition \ref{s-map} in the next subsection.

\subsection{The case of quantum operation}

The simplest example of quantum operation (completely positive trace
non-increasing linear map) from $\mathfrak{T}(\mathcal{H})$ to
$\mathfrak{T}(\mathcal{H}')$ is the map $(\cdot)\mapsto
V(\cdot)V^{*}$, where $V$ is a linear contraction from $\mathcal{H}$
to $\mathcal{H}'$. Theorem \ref{bound-cont} implies the following
result. \vspace{5pt}
\begin{property}\label{s-map}
\textit{Let $\,V$ be a linear operator from $\mathcal{H}$ to
$\mathcal{H}'$. The function
$\mathfrak{S}(\mathcal{H})\ni\rho\mapsto H(V\rho V^{*})$ is
continuous if and only if the operator $V$ is compact and has such
sequence $\{\nu_{i}\}$ of singular values (eigenvalues of
$\sqrt{V^{*}V}$) that
$\;\sum_{i=1}^{+\infty}e^{-\lambda/\nu^{2}_{i}}<+\infty$ for some
$\lambda>0$.} \emph{If this condition holds then}
$$
\sup_{\rho\in\mathfrak{S}(\mathcal{H})}H(V\rho V^{*})=\lambda^{*}(V)
$$
\emph{where $\lambda^{*}(V)$ is either the unique solution of the
equation $\,\sum_{i=1}^{+\infty}e^{-\lambda/\nu^{2}_{i}}=1$ if it
exists or equal to
$\,\mathrm{g}(\{\nu^{-2}_{i}\})=\inf\{\lambda>0\,|\,\sum_{i=1}^{+\infty}e^{-\lambda/\nu^{2}_{i}}<+\infty\}$
otherwise.}\footnote{The equation
$\,\sum_{i=1}^{+\infty}e^{-\lambda/\nu^{2}_{i}}=1$ has no solution
if and only if
$\,\sum_{i=1}^{+\infty}e^{-\mathrm{g}(\{\nu^{-2}_{i}\})/\nu^{2}_{i}}<1$.}
\end{property}\vspace{5pt}

In what follows we will use the  parameter $\lambda^{*}(V)$ defined
in Proposition \ref{s-map} for arbitrary operator
$V\in\mathfrak{B}(\mathcal{H})$ assuming that
$\lambda^{*}(V)=+\infty$ if the operator $V$  either is not compact
or has such sequence of singular values $\{\nu_{i}\}$ that
$\;\sum_{i=1}^{+\infty}e^{-\lambda/\nu^{2}_{i}}=+\infty$ for all
$\lambda>0$.\vspace{5pt}

\textbf{Proof.} We can assume that $\mathcal{H}=\mathcal{H}'$,
$V=|V|$, $\|V\|\leq 1$ and $\mathrm{Ker}V=\{0\}$.

Let $V=\sum_{i=1}^{+\infty}\nu_{i}|i\rangle\langle i|$. If
$\sum_{i=1}^{+\infty}e^{-\lambda/\nu^{2}_{i}}<+\infty$ for
$\lambda>0$ then property $\mathrm{(iv)}$ in Theorem
\ref{bound-cont} holds with the basis
$\{|i\rangle\}_{i=1}^{+\infty}$ and the sequence $\{h_{i}=\lambda
\nu_{i}^{-2}\}_{i=1}^{+\infty}$ (since in this case
$\Phi^{*}(\cdot)=V(\cdot)V$ and hence $\Phi^{*}(|i\rangle\langle
i|)=\nu^{2}_{i}|i\rangle\langle i|$).

The assertion concerning the supremum of the function $\rho\mapsto
H(V\rho V)$ is easily derived from Lemma \ref{simple+} in the
Appendix by using relation (\ref{H-q-c}).

Suppose the function $\rho\mapsto H(V\rho V)$ is continuous on the
set $\mathfrak{S}(\mathcal{H})$. Then the entropy is bounded on the
convex set $\{V\rho V\,|\,\rho\in\mathfrak{S}(\mathcal{H})\}$ and
hence this set is relatively compact by corollary 5 in \cite{Sh-4}
(used with the construction from the proof of Lemma \ref{H-oper}).
Thus the operator $V$ is compact (since otherwise there exists a
sequence of unit vectors $\{|\varphi_{n}\rangle\}$ such that the
sequence $\{V|\varphi_{n}\rangle\}$ is not relatively compact).
Lemma \ref{H-oper} implies existence of an operator
$T\in\mathfrak{T}_{1}(\mathcal{H})$ such that
$\sup_{\rho\in\mathfrak{S}(\mathcal{H})}\mathrm{Tr}V\rho V(-\log
T)<+\infty$ and $UT=TU$ for arbitrary unitary $U$ commuting with the
operator $V$. It follows from the last property of the operator $T$
that this operator is diagonizable in the basis $\{|i\rangle\}$,
t.i. $T=\sum_{i=1}^{+\infty}\tau_{i}|i\rangle\langle i|$, where
$\{\tau_{i}\}_{i=1}^{+\infty}$ is a sequence of nonnegative numbers
such that $\sum_{i=1}^{+\infty}\tau_{i}\leq 1$. Thus we have
$$
\sup_{\rho\in\mathfrak{S}(\mathcal{H})}\mathrm{Tr}V\rho V(-\log
T)=\sup_{\rho\in\mathfrak{S}(\mathcal{H})}\sum_{i=1}^{+\infty}\langle
i|\rho|i\rangle \nu^{2}_{i}(-\log\tau_{i})=\lambda<+\infty
$$
and hence $\nu^{2}_{i}(-\log\tau_{i})\leq\lambda$ for all $i$. This
implies $\lambda^{*}(V)<+\infty$. $\square$\vspace{5pt}

An arbitrary quantum operation
$\Phi:\mathfrak{T}(\mathcal{H})\rightarrow\mathfrak{T}(\mathcal{H}')$
has Kraus representation (\ref{Kraus-rep}) determined by a set
$\{V_{i}\}_{i=1}^{+\infty}$ of linear operators from $\mathcal{H}$
into $\mathcal{H}'$ such that
$\sum_{i=1}^{+\infty}V^{*}_{i}V_{i}\leq I_{\mathcal{H}}$. The
following proposition contains the sufficient conditions for
continuity of the output entropy of a quantum operation $\Phi$
expressed in terms of the set $\{V_{i}\}_{i=1}^{+\infty}$ of its
Kraus operators.\vspace{5pt}

\begin{property}\label{CE-conditions}
\textit{Let $\,\Phi$ be a quantum operation in $\,\mathfrak{F}_{\leq
1}(\mathcal{H},\mathcal{H}')$ and $\,\{V_{i}\}_{i\in I}$ be the
corresponding set of Kraus operators. Let
$d_{i}=\mathrm{rank}V_{i}\leq+\infty$.}\vspace{5pt}

A) \textit{If the set $\,I$ is finite then the operation $\,\Phi$
has continuous output entropy if and only if
\begin{equation}\label{n-s-cond}
\lambda^{*}\left(V_{i}\right)<+\infty\quad \forall i\in I,
\end{equation}
which in this case is equivalent to}
\begin{equation}\label{n-s-cond+}
\lambda^{*}\left(\sqrt{\textstyle\sum_{i\in
I}V^{*}_{i}V_{i}}\right)<+\infty. \footnote{In contrast to the set
$\{V_{i}\}_{i\in I}$ the operator $\sum_{i\in
I}V^{*}_{i}V_{i}=\Phi^{*}(I_{\mathcal{H}'})$ is uniquely determined
by the operation $\Phi$.}
\end{equation}

\textit{In general case (\ref{n-s-cond}) is a necessary condition of
continuity of the output entropy of the operation $\,\Phi$ (in
contrast to (\ref{n-s-cond+})).}\vspace{5pt}

B) \textit{If $\,I=\mathbb{N}$ then the operation $\,\Phi$ has
continuous output entropy if one of the following conditions is
valid:}
\begin{enumerate}[a)]

\item \textit{$d_{i}<+\infty$ for all $i$ and
there exists a sequence $\{h_{i}\}_{i=1}^{+\infty}$ of nonnegative
numbers such that}
$$
\|\sum_{i=1}^{+\infty}h_{i}V^{*}_{i}V_{i}\|<+\infty\quad\textit{and}\quad
\sum_{i=1}^{+\infty}d_{i}e^{-h_{i}}<+\infty;
$$
\item \textit{$H\left(\left\{\mathrm{Tr}V_{i}\rho
V_{i}^{*}\right\}^{+\infty}_{i=1}\right)<+\infty$ for all
$\rho\in\mathfrak{S}(\mathcal{H})$ \footnote{The sense of this
condition and its "variations" are considered in Proposition
\ref{CE-conditions-c} below.} and condition (\ref{n-s-cond+})
holds.}
\end{enumerate}
\vspace{5pt} \emph{Condition (\ref{n-s-cond+}) in $\,\mathrm{b)}$
can be replaced by the condition of finiteness of one of the series
$\,\sum_{i=1}^{+\infty}\lambda^{*}(V_{i})$ and
$\,\sum_{i=1}^{+\infty}\log d_{i}\|V_{i}\|^{2}$.}\vspace{5pt}

\textit{If the sequence $\,\{V_{i}\}_{i=1}^{+\infty}$ consists of
scalar multiples of mutually orthogonal projectors then
$\,\mathrm{a})$ is a necessary condition of continuity of the output
entropy of the operation $\,\Phi$.}\vspace{5pt}

\textit{If $\;\mathrm{Ran}V_{i}\perp\mathrm{Ran}V_{j}$ for all
$i\neq j$ then $\,\mathrm{b)}$ is a necessary condition of
continuity of the output entropy of the operation $\,\Phi$.}
\end{property}\vspace{5pt}

\textbf{Proof.} A) This directly follows from Corollary
\ref{bound-cont-c+}, Proposition \ref{s-map} and Corollary
\ref{basic-cont-cond+c} in Section 5 below (since
$\sum_{i=1}^{+\infty}V^{*}_{i}V_{i}=\Phi^{*}(I_{\mathcal{H}'})$).\vspace{5pt}

B) Suppose condition a) holds. Let
$\mathcal{K}=\bigoplus_{i=1}^{+\infty}\mathrm{Ran}V_{i}$ and $U_{i}$
be a partial isometry from $\mathcal{H}'$ into $\mathcal{K}$ such
that $U_{i}^{*}U_{i}$ is the projector onto
$\mathrm{Ran}V_{i}\subset\mathcal{H}'$ and $U_{i}U_{i}^{*}$ is the
projector onto $\mathrm{Ran}V_{i}\subset\mathcal{K}$. Consider the
quantum operation $\widehat{\Phi}$ in $\,\mathfrak{F}_{\leq
1}(\mathcal{H},\mathcal{K})$ defined by the sequence of Kraus
operators $\,\{\widehat{V}_{i}=U_{i}V_{i}\}_{i=1}^{+\infty}$.

We have
$\mathrm{Ran}\widehat{V}_{i}\perp\mathrm{Ran}\widehat{V}_{j}$ for
all $i\neq j$. Let $P_{i}$ be the $d_{i}$-rank projector onto the
subspace $\mathrm{Ran}\widehat{V}_{i}$. Consider the
$\mathfrak{H}$-operator $H=\sum_{i=1}^{+\infty}h_{i}P_{i}$. The
condition $\sum_{i=1}^{+\infty}d_{i}e^{-h_{i}}<+\infty$ means that
$\mathrm{g}(H)<+\infty$. The condition
$\|\sum_{i=1}^{+\infty}h_{i}V^{*}_{i}V_{i}\|=h<+\infty$ implies
$$
\mathrm{Tr}H\widehat{\Phi}(\rho)=\sum_{i=1}^{+\infty}h_{i}\mathrm{Tr}P_{i}\widehat{V}_{i}\rho
\widehat{V}_{i}^{*}=\mathrm{Tr}\sum_{i=1}^{+\infty}h_{i}V_{i}^{*}V_{i}\rho\leq
h,\quad \forall\rho\in\mathfrak{S}(\mathcal{H}).
$$
By Proposition \ref{H-cont-cond}A the quantum entropy is bounded on
the set $\widehat{\Phi}(\mathfrak{S}(\mathcal{H}))$. Since
$$
H(\widehat{\Phi}(\rho))=\sum_{i=1}^{+\infty}H(V_{i}\rho
V_{i}^{*})+H\left(\left\{\mathrm{Tr}V_{i}\rho
V_{i}^{*}\right\}_{i=1}^{+\infty}\right)
$$
inequality (\ref{w-k-ineq+}) implies boundedness of the function
$\rho\mapsto H(\Phi(\rho))$. By Theorem \ref{bound-cont} this
function is continuous. \vspace{5pt}

If condition b) holds then Proposition \ref{CE-conditions-c} below
and Corollary \ref{basic-cont-cond+c} in Section 5 imply continuity
of the output entropy of the operation $\Phi$. Possibility to
replace condition (\ref{n-s-cond+}) in b) by one of the conditions
$\sum_{i=1}^{+\infty}\lambda^{*}(V_{i})<+\infty$ and
$\sum_{i=1}^{+\infty}\log d_{i}\|V_{i}\|^{2}<+\infty$ follows from
Corollary \ref{bound-cont-c+}, since each of these conditions
implies finiteness of the series $\sum_{i=1}^{+\infty}H(V_{i}\rho
V_{i}^{*})$ for any $\rho$ in
$\mathfrak{S}(\mathcal{H})$.\vspace{5pt}

To prove the assertion concerning necessity of condition a) assume
that $\Phi(\cdot)=\sum_{i=1}^{+\infty}c_{i}P_{i}(\cdot)P_{i}$, where
$\{P_{i}\}$ is a sequence of mutually orthogonal projectors. Lemma
\ref{H-oper} implies existence of a trace class operator of the form
$T=\sum_{i=1}^{+\infty}\lambda_{i}P_{i}$ such that
$$
\sup_{\rho\in\mathfrak{S}(\mathcal{H})}\mathrm{Tr}\Phi(\rho)(-\log
T)=\sup_{\rho\in\mathfrak{S}(\mathcal{H})}\mathrm{Tr}\sum_{i=1}^{+\infty}c_{i}(-\log\lambda_{i})P_{i}\rho<+\infty.
$$
Since $\mathrm{Tr}T=\sum_{i=1}^{+\infty}d_{i}\lambda_{i}$, condition
a) holds with the sequence $\{h_{i}=-\log\lambda_{i}\}$.

To prove the assertion concerning necessity of condition b) it is
sufficient to note that the condition
$\;\mathrm{Ran}V_{i}\perp\mathrm{Ran}V_{j}$ for all $i\neq j$
implies
$$
H(V\rho V^{*})\leq H(\Phi(\rho))=\sum_{i=1}^{+\infty}H(V_{i}\rho
V_{i}^{*})+H\left(\left\{\mathrm{Tr}V_{i}\rho
V_{i}^{*}\right\}_{i=1}^{+\infty}\right)\quad \forall
\rho\in\mathfrak{S}(\mathcal{H}),
$$
where $V$ is the Stinespring contraction of the operation $\Phi$
defined via the set $\{V_{i}\}_{i=1}^{+\infty}$ (see the proof of
Proposition \ref{CE-conditions-c} below), and to apply Theorem
\ref{bound-cont} and Proposition \ref{s-map} (by using
$V^{*}V=\sum_{i=1}^{+\infty}V_{i}^{*}V_{i}$). $\square$\vspace{5pt}

\begin{example}\label{cont-cond+++r}
Let $\{V_{i}\}_{i=1}^{+\infty}$ be a sequence of finite rank
operators in $\mathfrak{B}(\mathcal{H})$ such that
$\sum_{i=1}^{+\infty}V^{*}_{i}V_{i}\leq I_{\mathcal{H}}$,
$\mathrm{Ran}V^{*}_{i}\perp\mathrm{Ran}V^{*}_{j}$ for all
sufficiently large\break $i\neq j$ and $\|V_{i}\|^{2}\leq
C\log^{-\alpha}(i)$ for all $i$, where $\alpha\geq0$ and $C>0$.
Since $V_{i}^{*}V_{i}\leq C\log^{-\alpha}(i) P_{i}$, where $P_{i}$
is the projector on the subspace $\mathrm{Ran}V^{*}_{i}$, condition
a) in Proposition \ref{CE-conditions}B holds for the operation
$\Phi_{\alpha}(\cdot)=\sum_{i=1}^{+\infty}V_{i}(\cdot)V^{*}_{i}$ for
all $\alpha\geq1$ provided the rate of increase of the sequence
$\{\mathrm{rank}V_{i}\}_{i}$ does not exceed the polynomial rate:
$\mathrm{rank}V_{i}\leq i^n$ for some natural $n$ and all
sufficiently large $i$ (this can be shown by using the sequence
$\{h_{i}=(n+2)\log(i)\}$). Hence the output entropy of the operation
$\Phi_{\alpha}$ is continuous in this case.

The last assertion of Proposition \ref{CE-conditions} shows that the
output entropy of the operation $\Phi_{\alpha}$ is not continuous if
$\alpha<1$ and $V_{i}=\sqrt{C\log^{-\alpha}(i)}P_{i}$ even for
bounded sequence $\{\mathrm{rank}V_{i}\}_{i}$. $\square$
\end{example}\vspace{5pt}

The following proposition contains the sufficient conditions for
continuity of the output entropy of the complementary operation
expressed in terms of the set $\{V_{i}\}_{i=1}^{+\infty}$ of Kraus
operators of the initial operation.\vspace{5pt}

\begin{property}\label{CE-conditions-c}
\textit{Let $\,\Phi$ be a quantum operation in $\,\mathfrak{F}_{\leq
1}(\mathcal{H},\mathcal{H}')$ and $\,\{V_{i}\}_{i=1}^{+\infty}$ be
the corresponding set of Kraus operators. The complementary
operation $\widetilde{\Phi}$ has continuous output entropy if one of
the following conditions \textup{(}related by $\,
\mathrm{c})\Rightarrow \mathrm{b})\Leftrightarrow
\mathrm{a})$\textup{)} is valid:}
\begin{enumerate}[a)]
\item \textit{$H\left(\left\{\mathrm{Tr}V_{i}\rho
V_{i}^{*}\right\}^{+\infty}_{i=1}\right)<+\infty$ for all
$\rho\in\mathfrak{S}(\mathcal{H})$;}
\item \textit{there exists a sequence $\{h_{i}\}_{i=1}^{+\infty}$ of nonnegative numbers
such that
$$
\|\sum_{i=1}^{+\infty}h_{i}V^{*}_{i}V_{i}\|<+\infty\quad\textit{and}\quad
\sum_{i=1}^{+\infty}e^{-h_{i}}<+\infty;
$$}
\item
$$
H\left(\left\{\|V_{i}\|^{2}\right\}_{i=1}^{+\infty}\right)<+\infty.
$$
\end{enumerate}
\textit{If $\;\mathrm{Ran}V_{i}\perp\mathrm{Ran}V_{j}$ for all
$i\neq j$ then $\,\mathrm{a})\Leftrightarrow \mathrm{b})$ is a
necessary condition of continuity of the output entropy of the
operation $\,\widetilde{\Phi}$.}
\end{property}\vspace{5pt}

\textbf{Proof.} Show first that a) implies continuity of the
function $\rho\mapsto H(\widetilde{\Phi}(\rho))$.

Let $\mathcal{H}''$ be a separable Hilbert space and $\{|i\rangle\}$
be an orthonormal basis in $\mathcal{H}''$. Then the operator
$V:\mathcal{H}\ni|\varphi\rangle\mapsto\sum_{i=1}^{+\infty}|V_{i}\varphi\rangle\otimes|i\rangle\in\mathcal{H}'\otimes\mathcal{H}''$
is the Stinespring contraction for the operation $\Phi$, t.i.
$$
\Phi(A)=\mathrm{Tr}_{\mathcal{H}''}VAV^{*},\quad
A\in\,\mathfrak{T}(\mathcal{H}).
$$
So we have
$$
\widetilde{\Phi}(A)=\mathrm{Tr}_{\mathcal{H}'}VAV^{*}=\sum_{i,j=1}^{+\infty}\mathrm{Tr}\left[V_{i}AV_{j}^{*}\right]|i\rangle\langle
j|,\quad A\in\,\mathfrak{T}(\mathcal{H}).
$$
By relation (\ref{H-q-c}) condition a) implies
$$
H(\widetilde{\Phi}(\rho))\leq H\left(\left\{\langle
i|\widetilde{\Phi}(\rho)|i\rangle\right\}_{i=1}^{+\infty}\right)=H\left(\left\{\mathrm{Tr}V_{i}\rho
V_{i}^{*}\right\}_{i=1}^{+\infty}\right)<+\infty,\quad
\rho\in\,\mathfrak{S}(\mathcal{H}).
$$
By Theorem \ref{bound-cont} the output entropy of the operation
$\widetilde{\Phi}$ is continuous.\vspace{5pt}

Since finiteness of the function
$\mathfrak{S}(\mathcal{H})\ni\rho\mapsto
H\left(\left\{\mathrm{Tr}V_{i}\rho
V_{i}^{*}\right\}^{+\infty}_{i=1}\right)$ implies its boundedness,
equivalence of conditions a) and b) can be proved by noting that the
last condition can be rewritten as follows
$$
\sup_{\rho\in\mathfrak{S}(\mathcal{H})}\sum_{i=1}^{+\infty}h_{i}\mathrm{Tr}V_{i}\rho
V_{i}^{*}<+\infty
$$
and by using the classical versions of Proposition
\ref{H-cont-cond}A and Lemma \ref{H-oper}.\vspace{5pt}

The implication $\mathrm{c)}\Rightarrow \mathrm{b)}$ is obvious.
\vspace{5pt}

If $\;\mathrm{Ran}V_{i}\perp\mathrm{Ran}V_{j}$ for all $i\neq j$
then
$$
\widetilde{\Phi}(\rho)=\sum_{i=1}^{+\infty}\mathrm{Tr}\left[V_{i}\rho
V_{i}^{*}\right]|i\rangle\langle i|,\quad
\rho\in\,\mathfrak{S}(\mathcal{H}),
$$
and hence the function in a) coincides with the output entropy of
the operation $\widetilde{\Phi}$. $\square$\vspace{5pt}

\begin{example}\label{cont-cond+++r+}
Let $\Phi_{\alpha}$ be the quantum operation described in Example
\ref{cont-cond+++r} with no rank restrictions on its Kraus
operators. By Proposition \ref{CE-conditions-c} the output entropy
of the operation $\widetilde{\Phi}_{\alpha}$ is continuous if
$\alpha\geq 1$ but it is not continuous  if $\alpha<1$ and
$V_{i}=\sqrt{C\log^{-\alpha}(i)}P_{i}$. $\square$
\end{example}\vspace{5pt}

\section{The PCE-property a positive linear map}

Continuity of the output entropy of a positive linear map on the
whole set of input states is a very strong requirement. In this
section we consider the substantially weaker property of a positive
linear map consisting in continuity of the output entropy on each
subset of input states on which the von Neumann entropy is
continuous.\vspace{5pt}

\subsection{The general case}

Let
$\,\Phi:\mathfrak{T}(\mathcal{H})\rightarrow\mathfrak{T}(\mathcal{H}')\,$
be a positive linear map. Since the output entropy $H_{\Phi}\doteq
H\circ\Phi$ of this map is a concave nonnegative lower
semicontinuous function on the cone $\mathfrak{T}_{+}(\mathcal{H})$,
we can apply the results of section 4 in \cite{Sh-11} to this
function as follows.

For each natural $k$ consider the concave function
\begin{equation}\label{out-ent-a}
H_{\Phi}^{k}(A)\doteq\sup_{\{\pi_{i},A_{i}\}\in\mathcal{P}^{\mathrm{a}}_{\{A\}}(\mathfrak{T}_{+}^{k}(\mathcal{H}))}
\sum_{i}\pi_{i}H_{\Phi}(A_{i})
\end{equation}
on the cone $\mathfrak{T}_{+}(\mathcal{H})$ (the supremum is over
all decompositions of the operator $A$ into countable convex
combination of operators of rank $\leq k$). By using
(\ref{H-fun-eq}) it is easy to see that the restriction of the above
function $H_{\Phi}^{k}$ to the set $\mathfrak{S}(\mathcal{H})$
coincides with the function $(\widehat{H_{\Phi}})^{\sigma}_{k}$
defined by (\ref{sigma-roof}) with $f=H_{\Phi}$ and that
$$
H_{\Phi}^{k}(\lambda A)=\lambda H_{\Phi}^{k}(A),\quad
A\in\mathfrak{T}_{+}(\mathcal{H}),\;\lambda\geq0.
$$

Hence by using propositions 1 and 3 in \cite{Sh-11} it is easy to
show that the function $H_{\Phi}^{k}$ is lower semicontinuous on the
cone $\mathfrak{T}_{+}(\mathcal{H})$ and that the monotonic sequence
$\{H_{\Phi}^{k}\}$ pointwise converges to the function $H_{\Phi}$.
Following \cite{Sh-11} we will call the function $H_{\Phi}^{k}$ the
\emph{$k$-order approximator} of the output entropy of the map
$\Phi$.

By using spectral decompositions one can prove uniform convergence
of the sequence $\{H_{\Phi}^{k}\}$ to the function $H_{\Phi}$ on
compact subsets of $\mathfrak{T}_{+}(\mathcal{H})$ on which the
quantum entropy is continuous.\vspace{5pt}
\begin{lemma}\label{uniform-c}
\textit{If the quantum entropy is continuous on a compact subset
$\mathcal{A}$ of $\,\mathfrak{T}_{+}(\mathcal{H})$ then}
\begin{equation}\label{uniform-c-exp}
\lim_{k\rightarrow+\infty}\sup_{A\in\mathcal{A},\Phi\in\mathfrak{L}_{\leq1}^{+}(\mathcal{H},\mathcal{H}')}
\left(H_{\Phi}(A)-H_{\Phi}^{k}(A)\right)=0.
\end{equation}
\end{lemma}\vspace{5pt}

\textbf{Proof.} We may assume that
$\mathcal{A}\subset\mathfrak{T}_{1}(\mathcal{H})$. Let
$\lambda_{i}^{k}(A)$ be the sum of the eigenvalues
$\lambda_{(i-1)k+1},...,\lambda_{ik}$ of the operator $A$ (arranged
in non-increasing order) and $P^{k}_{i}$ be the spectral projector
of this operator corresponding to the above set of eigenvalues.
Since the ensemble $\{\pi^{k}_{i}, (\pi^{k}_{i})^{-1}P^{k}_{i}A\}$,
where $\pi^{k}_{i}=\|A\|^{-1}_{1}\lambda^{k}_{i}(A)$, lies in
$\mathcal{P}^{\mathrm{a}}_{\{A\}}(\mathfrak{T}^{k}_{+}(\mathcal{H}))$,
by using inequalities (\ref{H-fun-ineq}) and (\ref{w-k-ineq+}) we
obtain
$$
\begin{array}{c}
\displaystyle H_{\Phi}(A)-H_{\Phi}^{k}(A)\leq
H(\Phi(A))-\sum_{i}\pi^{k}_{i}H\left(\Phi\left((\pi^{k}_{i})^{-1}P^{k}_{i}A\right)\right)
\\
\displaystyle =H(\Phi(A))-\sum_{i}H\left(\Phi(P^{k}_{i}A)\right)
\leq H\left(\{\mathrm{Tr}\Phi(P^{k}_{i}A)\}\right)\leq
H\left(\{\lambda_{i}^{k}(A)\}\right)
\end{array}
$$
for arbitrary map $\Phi$ in
$\mathfrak{L}_{\leq1}^{+}(\mathcal{H},\mathcal{H}')$. Hence the
assertion of the lemma follows from lemma 9 in \cite{Sh-11}, which
implies
$$
\lim_{k\rightarrow+\infty}\sup_{A\in\mathcal{A}}\widetilde{\Delta}_{k}(A)
=\lim_{k\rightarrow+\infty}\sup_{A\in\mathcal{A}}H\left(\{\lambda_{i}^{k}(A)\}\right)=0.\quad
\square
$$

Note that concavity of the function $\eta(x)=-x\log x$ implies the
inequality
$$
H_{\Phi}(A)-H_{\Phi}^{k}(A)\leq\inf_{\{\pi_{i},A_{i}\}\in\mathcal{P}^{\mathrm{a}}_{\{A\}}(\mathfrak{T}^{k}_{+}(\mathcal{H}))}
\sum_{i}\pi_{i}H(\Phi(A_{i})\|\Phi(A)),\quad
A\in\mathfrak{T}_{+}(\mathcal{H}),
$$
showing that (\ref{uniform-c-exp}) holds for arbitrary subset
$\mathcal{A}$ of $\mathfrak{T}_{+}(\mathcal{H})$ having the
UA\nobreakdash-\hspace{0pt}property (not necessarily compact)
provided that the set
$\mathfrak{L}_{\leq1}^{+}(\mathcal{H},\mathcal{H}')$ is replaced by
the set $\mathfrak{F}_{\leq 1}(\mathcal{H},\mathcal{H}')$ of all
quantum operations (or by any other subset of positive linear maps
for which monotonicity of the relative entropy holds). \vspace{5pt}

\begin{remark}\label{uniform-c+} Since
the function $A\mapsto H^{k}_{\Phi}(A)$ is lower semicontinuous on
the set $\mathfrak{T}_{+}(\mathcal{H})$ for each $k$, the
generalized Dini's lemma\footnote{The condition of continuity of
functions of the increasing sequence in the standard Dini's lemma
can be replaced by the condition of their lower semicontinuity
(provided that the condition of continuity of the limit function is
valid).} shows that continuity of the function $A\mapsto
H_{\Phi}(A)$ on a compact subset $\mathcal{A}$ of
$\,\mathfrak{T}_{+}(\mathcal{H})$ implies
$$
\lim_{k\rightarrow+\infty}\sup_{A\in\mathcal{A}}
\left(H_{\Phi}(A)-H_{\Phi}^{k}(A)\right)=0.
$$
The converse implication obviously holds if the function $A\mapsto
H^{k}_{\Phi}(A)$ is continuous on the set $\mathcal{A}$ for each
$k$. $\square$
\end{remark}\vspace{5pt}

The above observations imply the following answer on the second
question stated in the Introduction.\vspace{5pt}

\begin{theorem}\label{PCE-property}
\textit{Let $\,\Phi$ be a map in
$\mathfrak{L}_{\leq1}^{+}(\mathcal{H},\mathcal{H}')$. The following
properties are equivalent:}
\begin{enumerate}[(i)]
    \item \textit{the function $A\mapsto H_{\Phi}(A)$ is continuous on
    $\,\mathfrak{T}^{1}_{+}(\mathcal{H})$;}\footnote{
    $\,\mathfrak{T}^{1}_{+}(\mathcal{H})$ is the set of 1-rank positive operators in $\mathcal{H}$.}
    \item \textit{the function $A\mapsto H^{k}_{\Phi}(A)$ is continuous on  $\,\mathfrak{T}_{+}(\mathcal{H})$ for each $k$;}
    \item \textit{the function $A\mapsto H_{\Phi}(A)$ is continuous on an arbitrary subset of $\,\mathfrak{T}_{+}(\mathcal{H})$ on which the quantum entropy
    is continuous.}
\end{enumerate}
\textit{Property $\,\mathrm{(i)}$ is equivalent to continuity and
boundedness of the function\break $\rho\mapsto H_{\Phi}(\rho)$ on
the set $\mathrm{extr}\mathfrak{S}(\mathcal{H})$ and hence it
follows from the UA\nobreakdash-\hspace{0pt}property of the set
$\,\Phi(\mathrm{extr}\mathfrak{S}(\mathcal{H}))$.}

\end{theorem}\vspace{5pt}
\textbf{Proof.} $\mathrm{(i)\Rightarrow(ii)}$ Show first that
$\mathrm{(i)}$ implies continuity of the function $A\mapsto
H_{\Phi}(A)$ on the set $\mathfrak{T}_{+}^{k}(\mathcal{H})$ for each
$k$. Suppose there exists a sequence
$\{A_{n}\}\subset\mathfrak{T}_{+}^{k}(\mathcal{H})$ converging to an
operator $A_{0}$ such that
\begin{equation}\label{dc}
\lim_{n\rightarrow+\infty}H_{\Phi}(A_{n})>H_{\Phi}(A_{0}).
\end{equation}
For each $n\in\mathbb{N}$ we have $A_{n}=\sum_{i=1}^{k}A_{i}^{n}$,
where $\{A_{i}^{n}\}_{i=1}^{k}$ is a subset of
$\mathfrak{T}_{+}^{1}(\mathcal{H})$. Since the set
$\{A_{n}\}_{n\geq0}$ is compact the compactness criterion for
subsets of $\mathfrak{T}_{+}(\mathcal{H})$ (see the Appendix in
\cite{Sh-11}) implies relative compactness of the sequence
$\{A_{i}^{n}\}_{n}$ for each $i=\overline{1,k}$. Hence we may
consider that there exists
$\lim_{n\rightarrow+\infty}A_{i}^{n}=A_{i}^{0}\in\mathfrak{T}_{+}^{1}(\mathcal{H})$
for each $i=\overline{1,k}$. It is clear that
$\sum_{i=1}^{k}A_{i}^{0}=A_{0}$. It follows from $\mathrm{(i)}$ that
$\lim_{n\rightarrow+\infty}H_{\Phi}(A_{i}^{n})=H_{\Phi}(A_{i}^{0})$.
Hence Lemma \ref{face} implies a contradiction to (\ref{dc}).

Continuity of the function $H_{\Phi}$ on the set
$\mathfrak{T}_{+}^{k}(\mathcal{H})$ implies its boundedness on the
set $\mathfrak{S}_{k}(\mathcal{H})$. By corollary 1 in \cite{Sh-11}
the function $H^{k}_{\Phi}$ is continuous and bounded on the set
$\mathfrak{S}(\mathcal{H})$ and hence it is continuous on the set
$\mathfrak{T}_{+}(\mathcal{H})$.\vspace{5pt}

$\mathrm{(ii)\Rightarrow(iii)}$  directly follows from Lemma
\ref{uniform-c}. $\mathrm{(iii)\Rightarrow(i)}$ is obvious.
\vspace{5pt}

The last assertion of the theorem follows from theorem 2A in
\cite{Sh-11} (since the UA\nobreakdash-\hspace{0pt}property of a
bounded set obviously implies boundedness of the quantum entropy on
this set). $\square$\vspace{5pt}

\begin{remark}\label{PCE-operation-r}
The main assertion of Theorem \ref{PCE-property} (the implication
$\mathrm{(i)\Rightarrow(iii)}$) is based on the special properties
of the von Neumann entropy, it can not be proved by using only such
general properties of  entropy-type functions as concavity, lower
semicontinuity and nonnegativity. The simplest example showing this
is given by the function $A\mapsto
R_{0}(\Phi(A))=\|\Phi(A)\|_{1}\log\mathrm{rank}(\Phi(A))$ -- the
output $0$\nobreakdash-\hspace{0pt}order Renyi entropy of the map
$\Phi$.\footnote{Indeed, if $\Phi(A)=\frac{1}{2}(A+UAU^{*})$, where
$U$ is an unitary having no eigenvectors, then
$R_{0}(\Phi(A))=\|\Phi(A)\|_{1}\log2$ for all
$A\in\mathfrak{T}^{1}_{+}(\mathcal{H})$ but it is easy to see that
the function $A\mapsto R_{0}(\Phi(A))$ is not continuous on the set
$\mathfrak{T}^{ 2}_{+}(\mathcal{H})$ on which
$R_{0}(A)=\|A\|_{1}\log2$.} The essential roles in the proof of
Theorem \ref{PCE-property} are played by the second inequality in
(\ref{H-fun-ineq}) and the part $"if"$ of the assertion of Lemma
\ref{face}. $\square$
\end{remark}\vspace{5pt}

If property $\mathrm{(iii)}$ in Theorem \ref{PCE-property} holds for
a positive linear map $\Phi$ then we may say roughly speaking that
this map "preserves continuity of the entropy". This motivates the
following definition.\vspace{5pt}

\begin{definition}\label{PCE-operation}
\textit{A positive linear map (correspondingly quantum operation or
quantum channel) $\Phi$, for which property $\mathrm{(iii)}$ in
Theorem \ref{PCE-property} holds, is called
PCE\nobreakdash-\hspace{0pt}map (correspondingly
PCE\nobreakdash-\hspace{0pt}operation or
PCE\nobreakdash-\hspace{0pt}channel).}
\end{definition}\vspace{5pt}

The simplest examples of PCE\nobreakdash-\hspace{0pt}maps are
completely positive linear maps with finite Kraus representations,
for which property $\mathrm{(i)}$ in Theorem \ref{PCE-property}
obviously holds.\vspace{5pt}

By the last assertion of Theorem \ref{PCE-property} to prove the
PCE\nobreakdash-\hspace{0pt}property of a map $\Phi$ it is
sufficient to show that
$$
\Phi(\mathrm{extr}\mathfrak{S}(\mathcal{H}))\subseteq\Lambda(\mathcal{A}),
$$
where $\Lambda$ is a finite composition of set-operations preserving
the UA\nobreakdash-\hspace{0pt}property (see proposition 4 in
\cite{Sh-11}) and $\mathcal{A}$ is a compact set on which the
entropy is continuous. This implies the following
PCE\nobreakdash-\hspace{0pt}condition.\vspace{5pt}

\begin{corollary}\label{PCE-property-sc} \textit{A map $\,\Phi$ in
$\mathfrak{L}_{\leq1}^{+}(\mathcal{H},\mathcal{H}')$ is a
PCE\nobreakdash-\hspace{0pt}map if there exist separable Hilbert
space $\mathcal{K}$, family
$\{A_{\psi}\}_{\psi\in\mathcal{H},\|\psi\|=1}$ of operators
belonging to a particular compact subset $\mathcal{A}$ of
$\,\mathfrak{T}_{+}(\mathcal{K})$, on which the quantum entropy is
continuous, and family
$\,\{V_{\psi}\}_{\psi\in\mathcal{H},\|\psi\|=1}$ of linear
contractions from $\mathcal{K}$ to $\mathcal{H}'$ such that
$\,\Phi(|\psi\rangle\langle\psi|)=V_{\psi}A_{\psi}V_{\psi}^{*}$ for
each unit vector $\psi$ in $\mathcal{H}$.}
\end{corollary}\vspace{5pt}

If $\Phi$ is a quantum operation having the Kraus representation
with $k$ nonzero summands then the condition of Corollary
\ref{PCE-property-sc} trivially holds (with
$k$\nobreakdash-\hspace{0pt}dimensional Hilbert space
$\mathcal{K}$). Nontrivial application of Corollary
\ref{PCE-property-sc} is the proof of the
PCE\nobreakdash-\hspace{0pt}property for the following family of
quantum channels.\vspace{5pt}

\begin{example}\label{PCE-property-e} Let $\mathcal{H}_{a}$ be the Hilbert space
$\mathcal{L}_{2}([-a,+a])$, where $a<+\infty$, and
$\{U_{t}\}_{t\in\mathbb{R}}$ be the group of unitary operators in
$\mathcal{H}_{a}$ defined as follows
$$
(U_{t}\varphi)(x)=e^{-\mathrm{i}tx}\varphi(x),\quad\forall\varphi\in\mathcal{H}_{a}.
$$
For given probability density function $p(t)$ consider the quantum
channel
$$
\Phi_{p}^{a}:\mathfrak{T}(\mathcal{H}_{a})\ni
A\mapsto\int_{-\infty}^{+\infty}U_{t}A
U_{t}^{*}p(t)dt\in\mathfrak{T}(\mathcal{H}_{a}).
$$
In \cite{Sh-7} it is shown that the condition of Corollary
\ref{PCE-property-sc} holds for the channel $\Phi_{p}^{a}$ (with
some set of unitary operators $\{V_{\psi}\}$) provided that the
differential entropy of the distribution $p(t)$ is finite and that
the function $p(t)$ is bounded and monotonic on $(-\infty,-b]$ and
on $[+b,+\infty)$ for sufficiently large $b$. $\square$
\end{example}\vspace{5pt}

If property $\mathrm{(iii)}$ in Theorem \ref{PCE-property} holds for
two positive maps then it obviously holds for their composition,
hence this theorem  implies the following result.\vspace{5pt}
\begin{corollary}\label{PCE-property-c}
\textit{If property $\mathrm{(i)}$ in Theorem \ref{PCE-property}
holds for positive linear bounded maps
$\,\Phi:\mathfrak{T}(\mathcal{H})\rightarrow\mathfrak{T}(\mathcal{H}')$
and
$\,\Psi:\mathfrak{T}(\mathcal{H}')\rightarrow\mathfrak{T}(\mathcal{H}'')$
then it holds for the map
$\,\Psi\circ\Phi:\mathfrak{T}(\mathcal{H})\rightarrow\mathfrak{T}(\mathcal{H}'')$.}
\end{corollary}\vspace{5pt}

In quantum information theory the notion of the Convex Closure of
the Output Entropy (CCoOE) of a quantum channel (considered as a
function on the set of input states of this channel) is widely used
\cite{AB,Sh-7}. By generalizing the proof of proposition 2 in
\cite{Sh-7} it is easy to show that property $\mathrm{(i)}$ in
Theorem \ref{PCE-property} of a positive linear map $\Phi$ is
equivalent to continuity and boundedness of the CCoOE of this map on
the set $\mathfrak{S}(\mathcal{H})$. Hence Corollary
\ref{PCE-property-c} shows that \textit{continuity and boundedness
of the CCoOE of positive linear bounded maps\break
$\,\Phi:\mathfrak{T}(\mathcal{H})\rightarrow\mathfrak{T}(\mathcal{H}')$
and
$\,\Psi:\mathfrak{T}(\mathcal{H}')\rightarrow\mathfrak{T}(\mathcal{H}'')$
imply continuity and boundedness of the CCoOE of the map
$\,\Psi\circ\Phi:\mathfrak{T}(\mathcal{H})\rightarrow\mathfrak{T}(\mathcal{H}'')$.}\vspace{5pt}

Lemma \ref{face} implies the PCE\nobreakdash-\hspace{0pt}analog of
Corollary \ref{bound-cont-c}.\vspace{5pt}
\begin{corollary}\label{PCE-property-convex}
\textit{Let $\,\Phi$ and $\,\Psi$ be  maps in
$\,\mathfrak{L}_{\leq1}^{+}(\mathcal{H},\mathcal{H}')$ and
$\lambda\in(0,1)$.  $\,\lambda\Phi+(1-\lambda)\Psi$ is a
PCE\nobreakdash-\hspace{0pt}map if and only if $\,\Phi$ and $\,\Psi$
are PCE\nobreakdash-\hspace{0pt}maps.}
\end{corollary}\vspace{5pt}

Thus the set of all PCE\nobreakdash-\hspace{0pt}maps is convex and
forms a face of the convex set
$\,\mathfrak{L}_{\leq1}^{+}(\mathcal{H},\mathcal{H}')$. This face
obviously contains the face of all maps in
$\,\mathfrak{L}_{\leq1}^{+}(\mathcal{H},\mathcal{H}')$ with
continuous output entropy.\vspace{5pt}

\subsection{The case of quantum operation}

Theorem \ref{PCE-property} implies the following
observation.\vspace{5pt}

\begin{property}\label{PCE-conditions}
\textit{Let $\,\Phi$ be a quantum operation in $\,\mathfrak{F}_{\leq
1}(\mathcal{H},\mathcal{H}')$ and $\,\widetilde{\Phi}$ be its
complementary operation. The following properties are equivalent:}
\begin{enumerate}[(i)]
\item \textit{$\,\Phi$ is a PCE\nobreakdash-\hspace{0pt}operation;}

\item \textit{$\,\widetilde{\Phi}$ is a PCE\nobreakdash-\hspace{0pt}operation.}
\end{enumerate}\vspace{5pt}
\textit{In terms of the set $\,\{V_{i}\}_{i=1}^{+\infty}$ of Kraus
operators of the operation $\,\Phi$ a sufficient condition for
$\;\mathrm{(i)-(ii)}\,$ can be expressed in one of the following
forms:}
\begin{enumerate}[a)]
\item \textit{the function
$\,\varphi\mapsto
H\left(\left\{\|V_{i}|\varphi\rangle\|^{2}\right\}_{i=1}^{+\infty}\right)\,$
is continuous on the space $\mathcal{H}$;}

\item \textit{one of the conditions in Proposition \ref{CE-conditions-c} holds for the sequence $\,\{V_{i}\}_{i=1}^{+\infty}$.}
\end{enumerate}\vspace{5pt}
\textit{If $\,\Phi$ is a PCE\nobreakdash-\hspace{0pt}operation and
$\;\mathrm{Ran}V_{i}\perp\mathrm{Ran}V_{j}$ for all $i\neq j$ then
$\,\mathrm{a)}$ holds. }

\end{property}\vspace{5pt}

\textbf{Proof.} By Theorem \ref{PCE-property} equivalence of
$\mathrm{(i)}$ and $\mathrm{(ii)}$ follows from coincidence of the
output entropies of the operations $\Phi$ and $\widetilde{\Phi}$ on
the set of pure states.

Sufficiency of condition a) follows from the corresponding assertion
of Corollary \ref{PCE-conditions+} in Section 6 below (with
$\overline{V}_{n}=\{V_{i}\}_{i=1}^{+\infty}$ for all $n$).

Sufficiency of condition b) follows from the first assertion of this
proposition (since continuity of the output entropy implies the
PCE\nobreakdash-\hspace{0pt}property).

Necessity of condition a) in the case
$\;\mathrm{Ran}V_{i}\perp\mathrm{Ran}V_{j}$ for all $i\neq j$ is
obvious since in this case
$$
H\left(\sum_{i=1}^{+\infty}V_{i}|\varphi\rangle\langle\varphi|V^{*}_{i}\right)
=
H\left(\left\{\|V_{i}|\varphi\rangle\|^{2}\right\}_{i=1}^{+\infty}\right).\;\square
$$

By Corollary \ref{tensor-pr} the tensor product of two quantum
operations with continuous output entropy is a quantum operation
with continuous output entropy. The
PCE\nobreakdash-\hspace{0pt}property is not preserved in general
with respect to tensor products. The simplest example showing this
is the quantum channel
$\mathrm{Id}_{\mathcal{H}}\otimes\widetilde{\mathrm{Id}}_{\mathcal{H}}$,
which is not a PCE\nobreakdash-\hspace{0pt}channel if
$\dim\mathcal{H}=+\infty$. By Corollary \ref{tensor-pr} and
Proposition \ref{PCE-conditions} $\,\Phi\otimes\Psi$ is a
PCE\nobreakdash-\hspace{0pt}operation if either the operations
$\Phi$ and $\Psi$ or the operations $\widetilde{\Phi}$ and
$\widetilde{\Psi}$ have continuous output entropy. The following
proposition contains several observations concerning the
PCE\nobreakdash-\hspace{0pt}property of the map $\Phi\otimes\Psi$.
\vspace{5pt}

\begin{property}\label{PCE-property-cp}
\textit{Let
$\,\Phi:\mathfrak{T}(\mathcal{H})\rightarrow\mathfrak{T}(\mathcal{H}')$
and
$\Psi:\mathfrak{T}(\mathcal{K})\rightarrow\mathfrak{T}(\mathcal{K}')$
be quantum operations.} \vspace{5pt}

A) \emph{If $\,\Psi$ is a finite-dimensional operation
\textup{(}$\max\{\dim\mathcal{K},\dim\mathcal{K}'\}<+\infty$\textup{)}
then $\,\Phi\otimes\Psi$ is a PCE\nobreakdash-\hspace{0pt}operation
if and only if $\,\Phi$ is a
PCE\nobreakdash-\hspace{0pt}operation.}\footnote{The condition
$\dim\mathcal{K}<+\infty$ is essential. This can be shown by the
example $\Phi=\mathrm{Id}_{\mathcal{H}}$ and
$\Psi=\widetilde{\mathrm{Id}}_{\mathcal{K}}$.} \vspace{5pt}

B) \emph{If $\,\dim\mathcal{H}=\dim\mathcal{K}=+\infty$ and $\,\Psi$
is a quantum channel such that its complementary channel
$\,\widetilde{\Psi}$ has finite output entropy then the following
properties are equivalent:}
\begin{enumerate}[(i)]
    \item  \emph{$H(\Phi\otimes\Psi(|\varphi\rangle\langle\varphi|))<+\infty$ for
    any unit vector $\varphi\in\mathcal{H}\otimes\mathcal{K}$;}
    \item  \emph{$H(A)<+\infty\,\Rightarrow\, H(\Phi\otimes\Psi(A))<+\infty$ for
    any operator $A\in\mathfrak{T}_{+}(\mathcal{H}\otimes\mathcal{K})$;}
    \item \emph{$\Phi\otimes\Psi$ is a
    PCE\nobreakdash-\hspace{0pt}operation;}
\end{enumerate}
\emph{If $\,\Phi$ is a quantum channel then the condition of
finiteness of the output entropy of the channel $\,\widetilde{\Psi}$
can be replaced by the condition}
$$
\min\left\{H_{\widetilde{\Phi}}(\mathrm{Tr}_{\mathcal{K}}|\varphi\rangle\langle\varphi|),H_{\widetilde{\Psi}}(\mathrm{Tr}_{\mathcal{H}}|\varphi\rangle\langle\varphi|)\right\}
<+\infty\quad\forall\varphi\in\mathcal{H}\otimes\mathcal{K}.
$$
\end{property}
\vspace{5pt}

\begin{remark}\label{PCE-property-cp-r}
By Theorem \ref{PCE-property} $\Phi\otimes\Psi$ is a
PCE\nobreakdash-\hspace{0pt}operation if and only if the function
$\varphi\mapsto H(\Phi\otimes\Psi(|\varphi\rangle\langle\varphi|))$
is continuous and bounded on the unit sphere of
$\mathcal{H}\otimes\mathcal{K}$. Proposition \ref{PCE-property-cp}
shows that continuity and boundedness of this function follows from
its finiteness (provided the condition of this proposition holds).
Proposition \ref{PCE-property-cp} also shows that the operation
$\Phi\otimes\Psi$ \emph{preserves continuity of the entropy} (the
PCE\nobreakdash-\hspace{0pt}property) if it \emph{preserves
finiteness of the entropy} (property $\mathrm{(ii)}$).
\end{remark} \vspace{5pt}

\textbf{Proof.} A) The PCE\nobreakdash-\hspace{0pt}property of the
operation $\Phi\otimes\Psi$ obviously implies the same property of
the operation $\Phi$. To prove the converse implication it is
sufficient to show that $\mathrm{Id}_{\mathcal{H}'}\otimes\Psi$ and
$\Phi\otimes\mathrm{Id}_{\mathcal{K}}$ are
PCE\nobreakdash-\hspace{0pt}operations.

The operation $\mathrm{Id}_{\mathcal{H}'}\otimes\Psi$ has the
PCE\nobreakdash-\hspace{0pt}property since it has a finite Kraus
representation. By Theorem \ref{PCE-property} the
PCE\nobreakdash-\hspace{0pt}property of the operation
$\Phi\otimes\mathrm{Id}_{\mathcal{K}}$ follows from continuity and
boundedness of the function
\begin{equation}\label{function}
\mathrm{extr}\mathfrak{S}(\mathcal{H}\otimes\mathcal{K})\ni\omega\mapsto
H_{\Phi\otimes\mathrm{Id}_{\mathcal{K}}}(\omega)=
H_{\widetilde{\Phi}\otimes\widetilde{\mathrm{Id}}_{\mathcal{K}}}(\omega)=H\left(\widetilde{\Phi}\left(\mathrm{Tr}_{\mathcal{K}}\omega\right)\right).
\end{equation}
Since the map $\omega\mapsto\mathrm{Tr}_{\mathcal{K}}\omega$ is a
 continuous surjection from
$\mathrm{extr}\mathfrak{S}(\mathcal{H}\otimes\mathcal{K})$ onto
$\mathfrak{S}_{k}(\mathcal{H})$, where $k=\dim\mathcal{K}$, the
function (\ref{function}) is continuous and bounded if  the function
$H_{\widetilde{\Phi}}$ is continuous and bounded on the set
$\mathfrak{S}_{k}(\mathcal{H})$. By Theorem \ref{PCE-property} and
Proposition \ref{PCE-conditions} this holds if (and only if) $\Phi$
is a PCE\nobreakdash-\hspace{0pt}operation.

B) By applying Corollary \ref{tensor-pr+} to the operation
$\widetilde{\Phi}$ and to the channel $\widetilde{\Psi}$ we obtain
that $\mathrm{(i)}$ implies continuity of the output entropy of the
operation $\widetilde{\Phi}\otimes\widetilde{\Psi}$ (since
$H(\Phi\otimes\Psi(|\varphi\rangle\langle\varphi|))=H(\widetilde{\Phi}\otimes\widetilde{\Psi}(|\varphi\rangle\langle\varphi|))$).
Hence $\mathrm{(iii)}$ follows from Proposition
\ref{PCE-conditions}. It is clear that
$\mathrm{(iii)\Rightarrow(ii)}$ and $\mathrm{(ii)\Rightarrow(i)}$.
$\square$ \vspace{5pt}

\begin{remark}\label{PCE-property-cp-r+}
Proposition \ref{PCE-property-cp} with
$\Psi=\mathrm{Id}_{\mathcal{K}}$ gives the conditions for the
PCE\nobreakdash-\hspace{0pt}property of the operation
$\Phi\otimes\mathrm{Id}_{\mathcal{K}}$. The
PCE\nobreakdash-\hspace{0pt}property for the tensor product of two
quantum operations $\Phi\in\mathfrak{F}_{\leq
1}(\mathcal{H},\mathcal{H}')$ and $\Psi\in\mathfrak{F}_{\leq
1}(\mathcal{K},\mathcal{K}')$ can be proved by showing that either
$\Phi\otimes\mathrm{Id}_{\mathcal{K}'}$ and
$\mathrm{Id}_{\mathcal{H}}\otimes\Psi$ or
$\Phi\otimes\mathrm{Id}_{\mathcal{K}}$ and
$\mathrm{Id}_{\mathcal{H}'}\otimes\Psi$ are
PCE\nobreakdash-\hspace{0pt}operations since
$\Phi\otimes\Psi=\mathrm{Id}_{\mathcal{H}'}\otimes\Psi\circ\Phi\otimes\mathrm{Id}_{\mathcal{K}}=
\Phi\otimes\mathrm{Id}_{\mathcal{K}'}\circ\mathrm{Id}_{\mathcal{H}}\otimes\Psi$.

Note that the PCE\nobreakdash-\hspace{0pt}property of the operation
$\Phi\otimes\Psi$ does not imply the
PCE\nobreakdash-\hspace{0pt}property of the above components. To
show this it is sufficient to consider the example
$\Phi=\widetilde{\mathrm{Id}}_{\mathcal{H}}$ and
$\Psi=\widetilde{\mathrm{Id}}_{\mathcal{K}}$ with
$\dim\mathcal{H}=\dim\mathcal{K}=+\infty$.
\end{remark}

\section{The output entropies of a pair\\ of complementary quantum operations}

The output entropies of two complementary quantum operations
coincide on the set of pure input states but they are  substantially
different functions on the whole space of input states. Nevertheless
the following relation between  continuity properties of these
functions can be established.\vspace{5pt}

\begin{property}\label{basic-cont-cond+}
\textit{Let
$\,\Phi:\mathfrak{T}(\mathcal{H})\rightarrow\mathfrak{T}(\mathcal{H}')$
be a quantum operation and $\;\widetilde{\Phi}\,$ be its
complementary operation. Let $\mathcal{A}$ be a subset of
$\,\mathfrak{T}_{+}(\mathcal{H})$ such that
$\min\{H_{\Phi}(A),H_{\widetilde{\Phi}}(A) \}<+\infty$ for all
$A\in\mathcal{A}$. Then continuity of the quantum entropy on the set
$\mathcal{A}$ implies continuity of the function $\,A\mapsto
\left(H_{\Phi}(A)-H_{\widetilde{\Phi}}(A)\right)$ on the set
$\,\mathcal{A}$.}
\end{property}\vspace{5pt}

The assertion of Proposition \ref{basic-cont-cond+} follows from the
more general assertion of Proposition \ref{g-basic-cont-cond+} in
Section 6 below.\vspace{5pt}

\begin{remark}\label{basic-cont-cond+r} If $\Phi$ is a quantum channel then
$\,H_{\Phi}(\rho)-H_{\widetilde{\Phi}}(\rho)\,$ is the coherent
information $I_{c}(\rho,\Phi)$ of this channel at a state $\rho$
\cite{H-SSQT, N&Ch}.
\end{remark}
\vspace{5pt}

\begin{corollary}\label{basic-cont-cond}
\textit{Let
$\,\Phi:\mathfrak{T}(\mathcal{H})\rightarrow\mathfrak{T}(\mathcal{H}')$
be a quantum channel and $\;\widetilde{\Phi}\,$ be its complementary
channel. If any two functions from the triple
$\{H,H_{\Phi},H_{\widetilde{\Phi}}\}$ are continuous on a particular
set $\mathcal{A}\subset\mathfrak{T}_{+}(\mathcal{H})$ then the third
one is also continuous on this set.}

\emph{The above assertion holds for a quantum operation
$\,\Phi:\mathfrak{T}(\mathcal{H})\rightarrow\mathfrak{T}(\mathcal{H}')$
if}
\begin{equation}\label{p-channel}
\lambda^{*}\left(\sqrt{I_{\mathcal{H}}-\Phi^{*}(I_{\mathcal{H}'})}\right)<+\infty.\footnote{The
parameter $\lambda^{*}(\cdot)$ is defined in Proposition \ref{s-map}
in Section 3.2.}
\end{equation}
\end{corollary}

\textbf{Proof.} By representations (\ref{Stinespring-rep}) and
(\ref{c-channel}) the first assertion of the corollary follows from
Proposition \ref{basic-cont-cond+} and Proposition
\ref{cont-cond-a-1} in the Appendix.\vspace{5pt}

The second assertion of the corollary is derived from the first one
by means of Lemma \ref{inv-imp} below since by representations
(\ref{Stinespring-rep}) and (\ref{c-channel}) we have
$\Phi=\Theta\circ\Lambda$  and
$\widetilde{\Phi}=\widetilde{\Theta}\circ\Lambda$, where
$\Theta(\cdot)=\mathrm{Tr}_{\mathcal{H}''}(\cdot)$ is a quantum
channel from $\mathfrak{T}(\mathcal{H}'\otimes\mathcal{H}'')$ into
$\mathfrak{T}(\mathcal{H}')$ and $\Lambda(\cdot)=V(\cdot)V^{*}$ is a
quantum operation from $\mathfrak{T}(\mathcal{H})$ into
$\mathfrak{T}(\mathcal{H}'\otimes\mathcal{H}'')$.
$\square$\vspace{5pt}

\begin{lemma}\label{inv-imp}
\emph{Let $V$ be a linear contraction from $\mathcal{H}$ into
$\mathcal{H}'$ such that \break
$\lambda^{*}\left(\sqrt{I_{\mathcal{H}}-V^{*}V}\right)<+\infty$.
Then continuity of the function $A\mapsto H(VAV^{*})$ on a
particular set $\mathcal{A}\subset\mathfrak{T}_{+}(\mathcal{H})$
implies\footnote{In fact, "is equivalent to", since the converse
implication holds for an arbitrary contraction $V$ by Theorem
\ref{PCE-property}.} continuity of the quantum entropy on this set.}
\end{lemma}\vspace{5pt}

\textbf{Proof.} Consider the quantum channel
$$
\mathfrak{T}(\mathcal{H})\ni
A\mapsto\Psi(A)=VAV^{*}\oplus\sqrt{I_{\mathcal{H}}-V^{*}V}A\sqrt{I_{\mathcal{H}}-V^{*}V}\in
\mathfrak{T}(\mathcal{H}'\oplus\mathcal{H}).
$$
By Proposition \ref{s-map} the function $A\mapsto
H(\sqrt{I_{\mathcal{H}}-V^{*}V}A\sqrt{I_{\mathcal{H}}-V^{*}V})$ is
continuous on the set $\mathfrak{T}_{+}(\mathcal{H})$. Hence the
function $A\mapsto H_{\Psi}(A)$ is continuous on the set
$\mathcal{A}$. Since the complementary channel $\widetilde{\Psi}$
has two dimensional output space the assertion of the lemma follows
from the first part of Corollary \ref{basic-cont-cond}.
$\square$\vspace{5pt}

\begin{corollary}\label{cont-cond+++p}
\textit{Let
$\,\Phi:\mathfrak{T}(\mathcal{H})\rightarrow\mathfrak{T}(\mathcal{H}')$
be a quantum channel (or  quantum operation satisfying condition
(\ref{p-channel})) such that the complementary channel (operation)
$\,\widetilde{\Phi}$ has finite output entropy.\footnote{The
sufficient conditions for this property expressed in terms of the
Kraus operators of the operation $\Phi$ is presented in Proposition
\ref{CE-conditions-c} in Section 3.2.} Then the function $A\mapsto
H_{\Phi}(A)$ is continuous on a set
$\mathcal{A}\subset\mathfrak{T}_{+}(\mathcal{H})$ if and only if the
quantum entropy is continuous on this set.}\vspace{5pt}
\end{corollary} 

The simplest class of quantum channels for which the condition of
Corollary \ref{cont-cond+++p} holds consists of channels with finite
Kraus representation, for which complementary channels have finite
dimensional output. \vspace{5pt}

\begin{corollary}\label{basic-cont-cond++}\footnote{This corollary can be considered as a generalization
of Proposition \ref{cont-cond-a-2},  since its application to the
channel $\Phi(A)=\!\sum_{i=1}^{+\infty}\langle
i|A|i\rangle|i\rangle\langle i|$ implies the assertion of that
proposition.} \textit{Let
$\;\Phi(\cdot)=\sum_{i=1}^{+\infty}V_{i}(\cdot)V_{i}^{*}$ be a
quantum channel (or quantum operation satisfying condition
(\ref{p-channel})) in
$\;\mathfrak{F}_{\leq1}(\mathcal{H},\mathcal{H}')$ such that
$\;\mathrm{Ran}V_{i}\perp\mathrm{Ran}V_{j}$ for all sufficiently
large $i\neq j$. Then continuity of the function  $A\mapsto
H_{\Phi}(A)$ on a set
$\,\mathcal{A}\subset\mathfrak{T}_{+}(\mathcal{H})$ implies
continuity of the quantum entropy on the set
$\mathcal{A}$.}\vspace{5pt}
\end{corollary}
\textbf{Proof.} Note first that the complementary operation
$\widetilde{\Phi}$ can be represented as follows
$$
\widetilde{\Phi}(A)=\sum_{i,j=1}^{+\infty}\mathrm{Tr}\left[V_{i}AV_{j}^{*}\right]|i\rangle
\langle j|,\quad A\in\,\mathfrak{T}(\mathcal{H}),
$$
where $\{|i\rangle\}$ is an orthonormal basis in the output space
$\mathcal{H}''$ of this operation.

Suppose $\;\mathrm{Ran}V_{i}\perp\mathrm{Ran}V_{j}$ for all $i,j\geq
n,i\neq j$. Let
$\Phi_{1}(\cdot)=\sum_{i=1}^{n-1}V_{i}(\cdot)V_{i}^{*}$ and
$\Phi_{2}(\cdot)=\sum_{i=n}^{+\infty}V_{i}(\cdot)V_{i}^{*}$ be
quantum operations. By Lemma \ref{face} continuity of the function
$\mathcal{A}\ni A\mapsto H(\Phi(A))=H(\Phi_{1}(A)+\Phi_{2}(A))$
implies continuity of the function $\mathcal{A}\ni A\mapsto
H(\Phi_{2}(A))$. By the condition
$$
H(\Phi_{2}(A))=\!\sum_{i=n}^{+\infty}H(V_{i}A
V_{i}^{*})+H\!\left(\left\{\mathrm{Tr}V_{i}AV_{i}^{*}\right\}_{i=n}^{+\infty}\right)=\!\sum_{i=n}^{+\infty}H(V_{i}A
V_{i}^{*})+H(\widetilde{\Phi}_{2}(A)),
$$
where
$\widetilde{\Phi}_{2}(A)=\sum_{i=n}^{+\infty}\mathrm{Tr}\left[V_{i}AV_{i}^{*}\right]|i\rangle
\langle i|$. Since the both terms in the right side of the above
expression are lower semicontinuous functions of $A$, continuity of
the function $\mathcal{A}\ni A\mapsto H(\Phi_{2}(A))$ implies
continuity of the function\break $\mathcal{A}\ni A\mapsto
H(\widetilde{\Phi}_{2}(A))$.

Consider the quantum channel
$\,\Pi(\cdot)=P(\cdot)P+(I_{\mathcal{H}''}-P)(\cdot)(I_{\mathcal{H}''}-P)\,$
in $\;\mathfrak{F}_{=1}(\mathcal{H}'',\mathcal{H}'')$, where
$P=\sum_{i=1}^{n-1}|i\rangle \langle i|$. Since
$$
\Pi(\widetilde{\Phi}(A))=\sum_{i,j=1}^{n-1}\mathrm{Tr}\left[V_{i}AV_{j}^{*}\right]|i\rangle
\langle j|+\widetilde{\Phi}_{2}(A),
$$
continuity of the function $\mathcal{A}\ni A\mapsto H(\Phi_{2}(A))$
implies continuity of the function $\mathcal{A}\ni A\mapsto
H(\Pi(\widetilde{\Phi}(A)))$ by Lemma \ref{face}, which is
equivalent to continuity of the function $\mathcal{A}\ni A\mapsto
H(\widetilde{\Phi}(A))$ by Corollary \ref{cont-cond+++p}. Hence the
function $\mathcal{A}\ni A\mapsto H(A)$ is continuous by Corollary
\ref{basic-cont-cond}. $\square$ \vspace{5pt}

\begin{remark}\label{corollaries}
The assertions of Corollaries \ref{basic-cont-cond},
\ref{cont-cond+++p} and \ref{basic-cont-cond++} do not hold for a
quantum operation $\Phi$ not satisfying condition (\ref{p-channel}).
This follows from the next corollary.
\end{remark}\vspace{5pt}

\begin{corollary}\label{basic-cont-cond+c}
\textit{Let $\,\Phi$ be a quantum operation in
$\;\mathfrak{F}_{\leq1}(\mathcal{H},\mathcal{H}')$ and}
$$
\mathfrak{T}_{\Phi}=\{A\in\mathfrak{T}_{1}(\mathcal{H})\,|\,\min\{H_{\Phi}(A),H_{\widetilde{\Phi}}(A)\}<+\infty\}.
$$

\textit{If
$\,\lambda^{*}\!\left(\!\sqrt{\Phi^{*}(I_{\mathcal{H}'})}\right)<+\infty$
then the function
$A\mapsto\left(H_{\Phi}(A)-H_{\widetilde{\Phi}}(A)\right)$ is
continuous on the set $\,\mathfrak{T}_{\Phi}$ and its absolute value
does not exceed
$\,\lambda^{*}\!\left(\!\sqrt{\Phi^{*}(I_{\mathcal{H}'})}\right)$.}\vspace{5pt}

\emph{If the functions $\rho\mapsto H_{\Phi}(\rho)$ and $\rho\mapsto
H_{\widetilde{\Phi}}(\rho)$ are continuous on the set
$\mathfrak{S}(\mathcal{H})$ then the operator
$\,\Phi^{*}(I_{\mathcal{H}'})$ satisfies the above condition.}
\end{corollary}\vspace{5pt}

\textbf{Proof.} Since $\Phi^{*}(I_{\mathcal{H}'})=V^{*}V$, where $V$
is the Stinespring contraction for the operation $\Phi$, the first
assertion of the corollary follows from Propositions \ref{s-map} and
\ref{basic-cont-cond+} while the second one -- from Proposition
\ref{s-map} and Proposition \ref{cont-cond-a-1} in the Appendix.
$\square$\vspace{5pt}

\section{The output entropy as a function\\ of a pair (map, input
state)}

In analysis of continuity of information characteristics of a
quantum channel as functions of a channel it is necessary to
consider the output entropy as a function of a pair (channel, input
state) and to explore continuity of this function with respect to
the Cartesian product (coordinate-wise) topology on the set of such
pairs \cite{Sh-H,L&S}. The same problem appears in study of quantum
channels by means of their approximation by quantum operations
\cite{Sh-H}.\vspace{5pt}

\subsection{The general continuity condition}

The central result of this subsection is the following
proposition.\vspace{5pt}

\begin{property}\label{PCE-property+}
\emph{Let $\,\{\Phi_{n}\}$ be a sequence of maps in
$\,\mathfrak{L}^{+}_{\leq1}(\mathcal{H},\mathcal{H}')$ converging to
a map $\,\Phi_{0}$. The following properties are equivalent:}
\begin{enumerate}[(i)]
    \item \emph{$\displaystyle\lim_{n\rightarrow+\infty}H_{\Phi_{n}}(A_{n})=H_{\Phi_{0}}(A_{0})<+\infty$
for any sequence
$\,\{A_{n}\}\!\subset\mathfrak{T}^{1}_{+}(\mathcal{H})$ converging
to an operator $A_{0}$;}\footnote{
    $\,\mathfrak{T}^{1}_{+}(\mathcal{H})$ is the set of 1-rank positive operators in $\mathcal{H}$.}
\item \emph{$\displaystyle\lim_{n\rightarrow+\infty}H(A_{n})=H(A_{0})<+\infty\;\,\Rightarrow\;\,
\lim_{n\rightarrow+\infty}H_{\Phi_{n}}(A_{n})=H_{\Phi_{0}}(A_{0})<+\infty
$ for any sequence $\,\{A_{n}\}\subset\mathfrak{T}_{+}(\mathcal{H})$
converging to an operator  $A_{0}$.}
\end{enumerate}
\end{property}
\vspace{5pt}

\textbf{Proof.} It suffice to show that
$\mathrm{(i)\Rightarrow(ii)}$. For given natural $k$ the obvious
modification of the first part of the proof of the implication
$\mathrm{(i)\Rightarrow(ii)}$ in Theorem \ref{PCE-property} implies
validity of $\mathrm{(ii)}$ for any sequence
$\{A_{n}\}\subset\mathfrak{T}_{+}^{k}(\mathcal{H})$. \vspace{5pt}

Suppose there exist $\varepsilon>0$ and a sequence
$\{\rho_{n}\}\subset\mathfrak{S}(\mathcal{H})$ converging to a state
$\rho_{0}$ such that
\begin{equation}\label{discont}
\lim_{n\rightarrow+\infty}H(\rho_{n})=H(\rho_{0})<+\infty\quad\textup{and}\quad
\lim_{n\rightarrow+\infty}H_{\Phi_{n}}(\rho_{n})>H_{\Phi_{0}}(\rho_{0})+3\varepsilon.
\end{equation}

It is easy to see that $\mathrm{(i)}$ implies
$\limsup_{n\rightarrow+\infty}\sup_{\rho\in\mathrm{extr}\mathfrak{S}(\mathcal{H})}H_{\Phi_{n}}(\rho)<+\infty$.
Hence we may consider that
\begin{equation}\label{bound}
\sup_{n>0}\sup_{\rho\in\mathrm{extr}\mathfrak{S}(\mathcal{H})}H_{\Phi_{n}}(\rho)<+\infty.
\end{equation}

By Lemma \ref{nv-2} in the Appendix used with (\ref{out-ent-est}),
(\ref{u-h-c}) and (\ref{bound}) we may assume existence of a
sequence
$\{A^{k}_{n}\}_{n}\subset\mathfrak{T}^{k}_{+}(\mathcal{H})$,
converging to an operator $A^{k}_{0}$ ($k\in\mathbb{N}$), such that
$B^{k}_{n}=\rho_{n}-A^{k}_{n}\geq 0$,
$$
H(\Phi_{n}(B^{k}_{n}))<\varepsilon\quad \textup{and} \quad
\gamma_{n}=\mathrm{Tr}\Phi_{n}(\rho_{n})h_{2}\left(\frac{\mathrm{Tr}\Phi_{n}(B^{k}_{n})}{\mathrm{Tr}\Phi_{n}(\rho_{n})}\right)<\varepsilon\quad
\forall n\geq0.
$$

By the remark at the begin of the proof
$|H(\Phi_{n}(A_{n}^{k}))-H(\Phi_{0}(A_{0}^{k}))|<\varepsilon$ for
all sufficiently large $n$. Since
$\Phi_{n}(\rho_{n})=\Phi_{n}(A^{k}_{n})+\Phi_{n}(B^{k}_{n})$ for
each $n\geq0$, by using inequality (\ref{H-fun-ineq}) we obtain
$$
H(\Phi_{n}(\rho_{n}))-H(\Phi_{0}(\rho_{0}))\leq
H(\Phi_{n}(A_{n}^{k}))-H(\Phi_{0}(A_{0}^{k}))+H(\Phi_{n}(B^{k}_{n}))+\gamma_{n}<3\varepsilon
$$
for all sufficiently large $n$, contradicting to (\ref{discont}).

By this contradiction and lower semicontinuity of the quantum
entropy property $\mathrm{(ii)}$ holds for an arbitrary sequence
$\{A_{n}\}\subset\mathfrak{S}(\mathcal{H})$. Its validity for an
arbitrary sequence $\{A_{n}\}\subset\mathfrak{T}_{+}(\mathcal{H})$
can be easily shown by using (\ref{out-ent-est}) and (\ref{bound}).
$\square$\vspace{5pt}

\begin{corollary}\label{PCE-property+c}
\textit{For an arbitrary subset $\mathcal{A}$ of
$\,\mathfrak{T}_{+}(\mathcal{H})$, on which the quantum entropy is
continuous, the function $\,(\Phi,A)\,\mapsto\, H_{\Phi}(A)$ is
continuous on the set $\,\mathfrak{F}^{k}_{\leq
1}(\mathcal{H},\mathcal{H}')\times\mathcal{A}$ for each natural
$k$.}\footnote{$\mathfrak{F}^{k}_{\leq 1}(\mathcal{H},\mathcal{H}')$
is the set of all quantum operations from
$\mathfrak{T}(\mathcal{H})$ to $\mathfrak{T}(\mathcal{H}')$ having
the Kraus representation consisting of $\,\leq k$ summands.}
\end{corollary}\vspace{5pt}

Let $\mathfrak{V}_{\leq1}(\mathcal{H},\mathcal{H}')$ be the set of
all sequences $\{V_{i}\}_{i=1}^{+\infty}$ of linear bounded
operators from $\mathcal{H}$ into $\mathcal{H}'$ such that
$\sum_{i=1}^{+\infty}V^{*}_{i}V_{i}\leq I_{\mathcal{H}}$ endowed
with the Cartesian product of the strong operator topology (the
topology of coordinate-wise strong operator convergence). In what
follows a sequence $\{V_{i}\}_{i=1}^{+\infty}$ in
$\mathfrak{V}_{\leq1}(\mathcal{H},\mathcal{H}')$ will be called
$\,\mathfrak{V}$\emph{\nobreakdash-\hspace{0pt}vector} and will be
denoted $\overline{V}$, the corresponding operator
$\sum_{i=1}^{+\infty}V^{*}_{i}V_{i}$ will be denoted
$\s\overline{V}\s$.\vspace{5pt}

Let $\{|i\rangle\}_{i=1}^{+\infty}$ be an orthonormal basis in a
separable Hilbert space $\mathcal{H}''$. Consider the maps
\begin{equation}\label{Phi-map}
\mathfrak{V}_{\leq1}(\mathcal{H},\mathcal{H}')\ni\overline{V}\mapsto
\Phi[\overline{V}](\cdot)=\sum_{i=1}^{+\infty}V_{i}(\cdot)V_{i}^{*}\in\mathfrak{F}_{\leq
1}(\mathcal{H},\mathcal{H}')
\end{equation}
and
\begin{equation}\label{Phi-c-map}
\mathfrak{V}_{\leq1}(\mathcal{H},\mathcal{H}')\ni\overline{V}\mapsto
\widetilde{\Phi}\left[\overline{V}\right](\cdot)=\sum_{i,j=1}^{+\infty}\mathrm{Tr}V_{i}(\cdot)V_{j}^{*}|i\rangle\langle
j|\in\mathfrak{F}_{\leq 1}(\mathcal{H},\mathcal{H}'').\!
\end{equation}

The following lemma shows, in particular, continuity of these maps
on the subset
$\mathfrak{V}_{=1}(\mathcal{H},\mathcal{H}')=\left\{\overline{V}\,|\,\s\overline{V}\s=I_{\mathcal{H}}\right\}$
of $\mathfrak{V}_{\leq1}(\mathcal{H},\mathcal{H}')$ corresponding to
the set of quantum channels.\vspace{5pt}

\begin{lemma}\label{phi-map-cont}
\emph{Let $\,\{\overline{V}_n\}$ be a sequence of
$\,\mathfrak{V}$\nobreakdash-\hspace{0pt}vectors in
$\,\mathfrak{V}_{\leq1}(\mathcal{H},\mathcal{H}')$ converging to a
$\,\mathfrak{V}$\nobreakdash-\hspace{0pt}vector $\,\overline{V}_0$.
The following properties are equivalent:}
\begin{enumerate}[(i)]
    \item \emph{the sequence $\left\{\s\overline{V}_{n}\s\right\}$ weakly converges to the operator $\,\s\overline{V}_{0}\s$;}
    \item \emph{the sequence $\left\{\Phi[\overline{V}_{n}]\right\}$ strongly converges to the map $\,\Phi[\overline{V}_{0}]$;}
    \item \emph{the sequence $\left\{\widetilde{\Phi}[\overline{V}_{n}]\right\}$ strongly converges
     to the map $\,\widetilde{\Phi}[\overline{V}_{0}]$.}
\end{enumerate}
\end{lemma}

\textbf{Proof.} $\mathrm{(i)}\Rightarrow\mathrm{(ii)}$ It suffice to
show that the sequence
$\left\{\Phi[\overline{V}_{n}](|\varphi\rangle\langle\varphi|)\right\}$
tends to the operator
$\Phi[\overline{V}_{0}](|\varphi\rangle\langle\varphi|)$ for
arbitrary unit vector $\varphi\in\mathcal{H}$. This can be done by
noting that
$\lim_{n\rightarrow+\infty}\langle\varphi|\s\overline{V}_{n}\s|\varphi\rangle=\langle\varphi|\s\overline{V}_{0}\s|\varphi\rangle$
implies
$$
\lim_{m\rightarrow+\infty}\sup_{n\geq0}\mathrm{Tr}\sum_{i>m}V^{n}_{i}\,|\varphi\rangle\langle\varphi|(V^{n}_{i})^{*}
=\lim_{m\rightarrow+\infty}\sup_{n\geq0}\sum_{i>m}\|V^{n}_{i}|\varphi\rangle\|^{2}=0\quad\forall\varphi\in\mathcal{H}.
$$

$\mathrm{(i)}\Rightarrow\mathrm{(iii)}$ It suffice to show that the
sequence
$\left\{\widetilde{\Phi}[\overline{V}_{n}](|\varphi\rangle\langle\varphi|)\right\}$
tends to the operator
$\widetilde{\Phi}[\overline{V}_{0}](|\varphi\rangle\langle\varphi|)$
for arbitrary unit vector $\varphi\in\mathcal{H}$. This can be done
by noting that $\mathrm{(i)}$ means
$\lim_{n\rightarrow+\infty}\mathrm{Tr}\,\widetilde{\Phi}[\overline{V}_{n}](|\varphi\rangle\langle\varphi|)
=\mathrm{Tr}\,\widetilde{\Phi}[\overline{V}_{0}](|\varphi\rangle\langle\varphi|)$
and that weak convergence of a sequence
$\{A_{n}\}\subset\mathfrak{T}_{+}(\mathcal{H}'')$ to an operator
$A_{0}\in\mathfrak{T}_{+}(\mathcal{H}'')$ such that
$\lim_{n\rightarrow+\infty}\mathrm{Tr}A_{n}=\mathrm{Tr}A_{0}$
implies its convergence in the trace norm \cite{D-A}.\vspace{5pt}

$\mathrm{(ii)}\Rightarrow\mathrm{(i)}$ and
$\mathrm{(iii)}\Rightarrow\mathrm{(i)}$ are obvious since
$$
\langle\varphi|\,\s\overline{V}\s\,|\varphi\rangle=\mathrm{Tr}\,\Phi[\overline{V}](|\varphi\rangle\langle\varphi|)=
\mathrm{Tr}\,\widetilde{\Phi}[\overline{V}](|\varphi\rangle\langle\varphi|)\quad\forall\varphi\in\mathcal{H}.\;\square
$$

Proposition \ref{PCE-property+} implies the following
observation.\vspace{5pt}

\begin{corollary}\label{PCE-conditions+}
\textit{Let $\,\{\overline{V}_{n}\}$ be a sequence of
$\,\mathfrak{V}$\nobreakdash-\hspace{0pt}vectors in
$\,\mathfrak{V}_{\leq1}(\mathcal{H},\mathcal{H}')$ converging to a
$\,\mathfrak{V}$\nobreakdash-\hspace{0pt}vector $\,\overline{V}_{0}$
such that property $\,\mathrm{(i)}$ in Lemma \ref{phi-map-cont}
holds and
\begin{equation}\label{h-function}
\lim_{n\rightarrow+\infty}H\left(\left\{\|V^{n}_{i}|\varphi_{n}\rangle\|^{2}\right\}_{i=1}^{+\infty}\right)=
H\left(\left\{\|V^{0}_{i}|\varphi_{0}\rangle\|^{2}\right\}_{i=1}^{+\infty}\right)<+\infty
\end{equation}
for any sequence $\{\varphi_{n}\}$ of vectors in $\mathcal{H}$
converging to a vector $\varphi_{0}$.} \emph{Then
$$
\lim_{n\rightarrow+\infty}H\left(\Phi[\overline{V}_{n}](A_{n})\right)=H\left(\Phi[\overline{V}_{0}](A_{0})\right)<+\infty
$$
and
$$
\lim_{n\rightarrow+\infty}H\left(\widetilde{\Phi}[\overline{V}_{n}](A_{n})\right)=H\left(\widetilde{\Phi}[\overline{V}_{0}](A_{0})\right)<+\infty
$$
for any sequence $\{A_{n}\}\subset\mathfrak{T}_{+}(\mathcal{H})$
converging to an operator $A_{0}$ such that
$\;\lim_{n\rightarrow+\infty}H(A_{n})=H(A_{0})<+\infty$.}\vspace{5pt}

\emph{The above requirements on the sequence
$\,\{\overline{V}_{n}\}$ can be replaced by one of the following
conditions:}
\begin{enumerate}[a)]

\item \textit{there exists a sequence $\{h_{i}\}_{i=1}^{+\infty}$ of nonnegative numbers such that
$$
\sup_{n\geq0}\|\sum_{i=1}^{+\infty}h_{i}(V_{i}^{n})^{*}V_{i}^{n}\|<+\infty\quad\textit{and}\quad
\sum_{i=1}^{+\infty}e^{-\lambda
h_{i}}<+\infty\;\,\textit{for}\;\,\textit{all}\;\, \lambda>0;
$$} 

\item \textit{property $\,\mathrm{(i)}$ in Lemma
\ref{phi-map-cont} holds for the sequence $\,\{\overline{V}_{n}\}$
and}
$$
\lim_{m\rightarrow+\infty}\sup_{n\geq0}H\left(\left\{\|V^{n}_{i}\|^{2}\right\}_{i>m}\right)=0.
$$
\end{enumerate}
\end{corollary}\vspace{5pt}

\textbf{Proof.} By using Proposition \ref{cont-cond-a-2} it is easy
to show that (\ref{h-function}) implies
$$
\lim_{n\rightarrow+\infty}H\left(\widetilde{\Phi}[\overline{V}_{n}](|\varphi_{n}\rangle\langle\varphi_{n}|)\right)
=H\left(\widetilde{\Phi}[\overline{V}_{0}](|\varphi_{0}\rangle\langle\varphi_{0}|)\right)<+\infty.
$$
for any sequence $\{\varphi_{n}\}$ of vectors in $\mathcal{H}$
converging to a vector $\varphi_{0}$. Hence the main assertion of
the corollary follows from Proposition \ref{PCE-property+} and Lemma
\ref{phi-map-cont} (since the output entropies of complementary
quantum operations coincide on the set of
$1$\nobreakdash-\hspace{0pt}rank operators).

Condition a) means that
$$
\sup_{n\geq0}\sup_{\varphi\in\mathcal{H},
\|\varphi\|\leq1}\sum_{i=1}^{+\infty}h_{i}\|V^{n}_{i}|\varphi\rangle\|^{2}<+\infty,
$$
which implies
$\lim_{m\rightarrow+\infty}\sup_{n\geq0}\sum_{i>m}^{+\infty}\|V^{n}_{i}|\varphi\rangle\|^{2}=0$
for each $\varphi\in\mathcal{H}$, showing that property
$\mathrm{(i)}$ in Lemma \ref{phi-map-cont} holds for the sequence
$\{\overline{V}_{n}\}$. By Proposition \ref{H-cont-cond} it also
implies (\ref{h-function}) for any sequence $\{\varphi_{n}\}$ of
vectors in $\mathcal{H}$ converging to a vector $\varphi_{0}$.

Condition b) implies
$$
\lim_{m\rightarrow+\infty}\sup_{n\geq0}\left[H\left(\left\{\|V^{n}_{i}|\varphi_{n}\rangle\|^{2}\right\}_{i=1}^{+\infty}\right)-
H\left(\left\{\|V^{n}_{i}|\varphi_{n}\rangle\|^{2}\right\}_{i=1}^{m}\right)\right]=0
$$
for any sequence $\{\varphi_{n}\}$ of vectors in $\mathcal{H}$
converging to a vector $\varphi_{0}$. Indeed, since weak convergence
of the sequence $\{\s\overline{V}_{n}\s\}$ to the operator
$\s\overline{V}_{0}\s$ implies
$\lim_{n\rightarrow+\infty}\langle\varphi_{n}|\s\overline{V}_{n}\s|\varphi_{n}\rangle=
\langle\varphi_{0}|\s\overline{V}_{0}\s|\varphi_{0}\rangle $ and
hence
$$
\lim_{m\rightarrow+\infty}\sup_{n\geq
0}\sum_{i>m}\|V^{n}_{i}|\varphi_{n}\rangle\|^{2}=0,
$$
the above assertion can be proved by using (\ref{H-fun-ineq}) and
(\ref{u-h-c}). $\square$\vspace{5pt}

\begin{example}\label{cont-cond+++r++}
Let $\,\{\overline{V}_{n}=\{V^{n}_{i}\}\}$ be a sequence of
$\,\mathfrak{V}$\nobreakdash-\hspace{0pt}vectors in
$\,\mathfrak{V}_{\leq1}(\mathcal{H},\mathcal{H}')$ converging to a
$\,\mathfrak{V}$\nobreakdash-\hspace{0pt}vector
$\,\overline{V}_{0}=\{V^{0}_{i}\}$ such that
$\mathrm{Ran}(V^{n}_{i})^{*}\perp(\mathrm{Ran}V^{n}_{j})^{*}$  and
$\|V^{n}_{i}\|^{2}\leq x_{i}\log^{-1}(i)$ for each $n$ and all
$i,j>m$,$\,i\neq j$, where $m\in\mathbb{N}$ and $\{x_{i}\}_{i>m}$ is
a given sequence of positive numbers converging to zero. Then
condition a) in Corollary \ref{PCE-conditions+} holds for the
sequence $\,\{\overline{V}_{n}\}$ with the sequence
$\{h_{i}\}_{i=1}^{+\infty}$, where $h_{i}=0\,$ if $\,i\leq m$ and
$h_{i}=x_{i}^{-1}\log(i)\,$ if $\,i>m$.
\end{example}\vspace{5pt}

\subsection{The continuity condition based on\\ the complementary relation}

By using the relation between complementary quantum operations the
following result can be established.\vspace{5pt}

\begin{property}\label{g-basic-cont-cond+}
\textit{Let $\,\{\Phi_{n}\}$ be a sequence of operations in
$\,\mathfrak{F}_{\leq 1}(\mathcal{H},\mathcal{H}')$ converging to an
operation $\,\Phi_{0}$ and $\,\{\widetilde{\Phi}_{n}\}$ be a
sequence of operations in $\,\mathfrak{F}_{\leq
1}(\mathcal{H},\mathcal{H}'')$ converging to an operation
$\,\widetilde{\Phi}_{0}$ such that $(\Phi_{n},\widetilde{\Phi}_{n})$
is a complementary pair for each $n=0,1,2..$. Let $\,\{A_{n}\}$ be a
sequence of operators in $\,\mathfrak{T}_{+}(\mathcal{H})$
converging to an operator $A_{0}$ such that
$$
\lim_{n\rightarrow+\infty}H(A_{n})=H(A_{0})<+\infty\;\;
\textit{and}\;\;\min\{H_{\Phi_{n}}(A_{n}),
H_{\widetilde{\Phi}_{n}}(A_{n})\}<+\infty,\;\; n\geq0.
$$ Then}
$$
\lim_{n\rightarrow+\infty}\left(H_{\Phi_{n}}(A_{n})-H_{\widetilde{\Phi}_{n}}(A_{n})\right)=
H_{\Phi_{0}}(A_{0})-H_{\widetilde{\Phi}_{0}}(A_{0})<+\infty.
$$
\end{property}\vspace{5pt}

\textbf{Proof.} Finiteness of the values $H_{\Phi_{n}}(A_{n})$ and
$H_{\widetilde{\Phi}_{n}}(A_{n})$ for each $n\geq0$ follows from the
definition of a complementary operation and inequality
(\ref{d-ineq}).

Let $\{A_{n}=\rho_{n}\}$ be a sequence of states in
$\mathfrak{S}(\mathcal{H})$ converging to a state $A_{0}=\rho_{0}$
such that
$\lim_{n\rightarrow+\infty}H(\rho_{n})=H(\rho_{0})<+\infty$.

Let
$a_{n}=H_{\Phi_{n}}(\rho_{n})-H_{\widetilde{\Phi}_{n}}(\rho_{n})$
for each $n\geq0$. By symmetry to prove that
$\lim_{n\rightarrow+\infty}a_{n}=a_{0}$ it is sufficient to show
that
\begin{equation}\label{rel-1}
 \liminf_{n\rightarrow+\infty}a_{n}\geq a_{0}.
\end{equation}

Let $\mathcal{K}$ be a separable Hilbert space and
$\{|\varphi_{n}\rangle\}$ be a sequence of unit vectors in
$\mathcal{H}\otimes\mathcal{K}$ converging to a vector
$|\varphi_{0}\rangle$ such that
$\mathrm{Tr}_{\mathcal{K}}|\varphi_{n}\rangle\langle\varphi_{n}|=\rho_{n}$
for each $n\geq0$. Finiteness of the values $H_{\Phi_{n}}(\rho_{n})$
and $H_{\widetilde{\Phi}_{n}}(\rho_{n})$ implies
$$
b_{n}=H(\Phi_{n}\otimes\mathrm{Id}_{\mathcal{K}}(|\varphi_{n}\rangle\langle\varphi_{n}|)\|
\Phi_{n}(\rho_{n})\otimes\rho_{n})=a_{n}+c_{n},
$$
where
$c_{n}=\mathrm{Tr}\,\Phi_{n}\otimes\mathrm{Id}_{\mathcal{K}}(|\varphi_{n}\rangle\langle\varphi_{n}|)\cdot
I_{\mathcal{H}'}\otimes(-\log\rho_{n})$.\vspace{5pt}

Since the sequence  $\{\Phi_{n}\otimes\mathrm{Id}_{\mathcal{K}}\}$
strongly converges to the operation
$\Phi_{0}\otimes\mathrm{Id}_{\mathcal{K}}$ we have
$\liminf_{n\rightarrow+\infty}b_{n}\geq b_{0}$ by lower
semicontinuity of the relative entropy. Hence to prove (\ref{rel-1})
we have to show that
\begin{equation}\label{rel-2}
\limsup_{n\rightarrow+\infty}c_{n}\leq c_{0}.
\end{equation}

For each $n$ consider the quantum channel
$\Psi_{n}=\Phi_{n}+\Delta_{n}$ in
$\mathfrak{F}_{=1}(\mathcal{H},\mathcal{H}')$, where
$\Delta_{n}(\cdot)=\sigma\mathrm{Tr}((I_{\mathcal{H}}-\Phi^{*}_{n}(I_{\mathcal{H}}))(\cdot))$
is a quantum operation defined by means of some state $\sigma$ in
$\mathfrak{S}(\mathcal{H}')$.  It is clear that the sequence
$\{\Delta_{n}\}$ strongly converges to the operation $\Delta_{0}$.
Since
$\,\lim_{n\rightarrow+\infty}H(\rho_{n})=H(\rho_{0})<+\infty\,$ and
$$
H(\rho_{n})=\mathrm{Tr}\,\Psi_{n}\otimes\mathrm{Id}_{\mathcal{K}}(|\varphi_{n}\rangle\langle\varphi_{n}|)\cdot
I_{\mathcal{H}'}\otimes(-\log\rho_{n})= c_{n}+d_{n},\quad
n=0,1,2,...,
$$
where
$d_{n}=\mathrm{Tr}\Delta_{n}\otimes\mathrm{Id}_{\mathcal{K}}(|\varphi_{n}\rangle\langle\varphi_{n}|)\cdot
I_{\mathcal{H}'}\otimes(-\log\rho_{n})$, to prove (\ref{rel-2}) it
suffice to show that
\begin{equation}\label{rel-3}
\liminf_{n\rightarrow+\infty}d_{n}\geq d_{0}.
\end{equation}
We have $d_{n}=\mathrm{Tr}B_{n}(-\log\rho_{n})$, where
$B_{n}=\mathrm{Tr}_{\mathcal{H}'}\Delta_{n}\otimes\mathrm{Id}_{\mathcal{K}}(|\varphi_{n}\rangle\langle\varphi_{n}|)$.
Since $B_{n}\leq
B_{n}+\mathrm{Tr}_{\mathcal{H}'}\Phi_{n}\otimes\mathrm{Id}_{\mathcal{K}}(|\varphi_{n}\rangle\langle\varphi_{n}|)=\rho_{n}$,
we have $H(B_{n})<+\infty$ and hence
$d_{n}=H(B_{n})+H(B_{n}\|\rho_{n})+\eta(\mathrm{Tr}B_{n})+\mathrm{Tr}B_{n}-1$.
Lower semicontinuity of the quantum entropy and of the relative
entropy implies (\ref{rel-3}).

Thus the assertion of the proposition is proved in the case
$\{A_{n}\}\subset\mathfrak{S}(\mathcal{H})$. The general assertion
is easily derived from the above observation by noting that for
arbitrary sequence $\{A_{n}\}$ converging to zero inequality
(\ref{d-ineq}) and the inequality $H(V^{n}A_{n}(V^{n})^{*})\leq
H(A_{n})$, where $V^{n}$ is the Stinespring contraction for the
operations $\Phi_{n}$ and $\widetilde{\Phi}_{n}$, show that
$$
\lim_{n\rightarrow+\infty}H(A_{n})=0\quad\Rightarrow\quad\lim_{n\rightarrow+\infty}\left(H_{\Phi_{n}}(A_{n})-H_{\widetilde{\Phi}_{n}}(A_{n})\right)=0.
\;\square
$$

\begin{corollary}\label{g-basic-cont-cond+c}
\textit{Let $\,\{\overline{V}_{n}\}$ be a sequence of
$\,\mathfrak{V}$\nobreakdash-\hspace{0pt}vectors in
$\,\mathfrak{V}_{\leq1}(\mathcal{H},\mathcal{H}')$ converging to a
$\,\mathfrak{V}$\nobreakdash-\hspace{0pt}vector $\,\overline{V}_{0}$
such that property $\,\mathrm{(i)}$ in Lemma \ref{phi-map-cont}
holds, and $\{A_{n}\}$ be a sequence of operators in
$\mathfrak{T}_{+}(\mathcal{H})$ converging to an operator $A_{0}$
such that $\;\lim_{n\rightarrow+\infty}H(A_{n})=H(A_{0})<+\infty$.
The following properties are equivalent:}
\begin{enumerate}[(i)]
    \item \emph{$\;\lim_{n\rightarrow+\infty}H\left(\Phi[\overline{V}_{n}](A_{n})\right)=H\left(\Phi[\overline{V}_{0}](A_{0})\right)<+\infty$;}
    \item \emph{$\;\lim_{n\rightarrow+\infty}H\left(\widetilde{\Phi}[\overline{V}_{n}](A_{n})\right)=
    H\left(\widetilde{\Phi}[\overline{V}_{0}](A_{0})\right)<+\infty$.}
\end{enumerate}
\emph{These properties hold if
    $$
    \lim_{n\rightarrow+\infty}H\left(\left\{\mathrm{Tr}V^{n}_{i}A_{n}(V^{n}_{i})^{*}\right\}_{i=1}^{+\infty}\right)=
    H\left(\left\{\mathrm{Tr}V^{0}_{i}A_{0}(V^{0}_{i})^{*}\right\}_{i=1}^{+\infty}\right)<+\infty.
    $$}
\end{corollary}

\textbf{Proof.} By Lemma \ref{phi-map-cont} the main assertion
directly follows from Proposition \ref{g-basic-cont-cond+}. The
second assertion is proved by using Proposition \ref{cont-cond-a-2}.
$\square$

\section{On continuity of the output entropy of\\ quantum
operations on a given set of states}

\subsection{The case of a single operation}

In analysis of quantum channels and operations the question of
continuity of their output entropy on a \emph{given} set of input
states naturally arises (see Section 8). By summarizing the results
of the previous sections we consider below the possibilities to
prove continuity of the output entropy of a quantum operation
$\Phi(\cdot)=\sum_{i=1}^{+\infty}V_{i}(\cdot)V^{*}_{i}$ on a given
set of states $\mathcal{A}$ in the two cases (distinguished by
accessible information about properties of this set).\vspace{5pt}

A) \emph{If the set $\mathcal{A}$ is arbitrary then the function
$\,\rho\mapsto H_{\Phi}(\rho)\,$ is continuous on this set provided
one of the following conditions is valid:}
\begin{enumerate}[1)]
    \item \emph{the quantum entropy is continuous on the set
    $\,\Phi(\mathcal{A})$;} \footnote{This condition is obviously sufficient but it is not necessary (see Remark \ref{on-cont}).}
    \item \emph{the operation $\,\Phi$ has continuous output entropy (sufficient conditions are  presented
in Section 3);}
    \item
    \emph{$\lambda^{*}\left(\sqrt{\Phi^{*}(I_{\mathcal{H}'})}\right)<+\infty$ and the function $\rho\mapsto
    H_{\widetilde{\Phi}}(\rho)$ is continuous on the set $\mathcal{A}$ (the last property can be verified by using Propositions \ref{CE-conditions-c} and \ref{comp-oper-c-c}).}
\end{enumerate}

B) \emph{If the von Neumann entropy is continuous on the set
$\mathcal{A}$ then the function $\,\rho\mapsto H_{\Phi}(\rho)\,$ is
continuous on this set provided one of the following conditions is
valid:}
\begin{enumerate}[1)]
     \item \emph{$\Phi$ is a PCE-operation (sufficient conditions are  presented in Section 4);}
    \item \emph{the  function $\rho\mapsto
    H_{\widetilde{\Phi}}(\rho)$ is continuous on the set $\mathcal{A}$ (Proposition \ref{comp-oper-c-c} below).}
\end{enumerate}

Conditions A-3 and B-2 follow respectively from Corollary
\ref{basic-cont-cond+c} and Proposition \ref{basic-cont-cond+}. To
verify  continuity of the function $\rho\mapsto
H_{\widetilde{\Phi}}(\rho)$ in these conditions one can use either
Proposition \ref{CE-conditions-c} or the following
proposition.\vspace{5pt}

\begin{property}\label{comp-oper-c-c}\ \textit{Let
$\;\Phi(\cdot)=\sum_{i=1}^{+\infty}V_{i}(\cdot)V^{*}_{i}$ be a
quantum operation and $\mathcal{A}$ be a subset of
$\,\mathfrak{S}(\mathcal{H})$. The  function $\rho\mapsto
H_{\widetilde{\Phi}}(\rho)$ is continuous on the set $\mathcal{A}$
if one of the following conditions \textup{(}related by $\,
\mathrm{b)}\Rightarrow \mathrm{a)}$\textup{)} is valid:}
\begin{enumerate}[a)]

\item \textit{the function $\rho\mapsto H\left(\left\{\mathrm{Tr}V_{i}\rho
V_{i}^{*}\right\}^{+\infty}_{i=1}\right)$ is continuous on the set
$\mathcal{A}$;}

\item \textit{there exists a sequence $\{h_{i}\}_{i=1}^{+\infty}$ of nonnegative numbers
such that
$$
\sup_{\rho\in\mathcal{A}}\mathrm{Tr}\sum_{i=1}^{+\infty}h_{i}V^{*}_{i}V_{i}\rho<+\infty\quad\textit{and}\quad
\sum_{i=1}^{+\infty}e^{-\lambda h_{i}}<+\infty\;\;
\textit{for}\;\;\textit{all}\;\;\lambda>0.
$$}
\end{enumerate}
\end{property}
\textbf{Proof.} Sufficiency of condition $\mathrm{a)}$ follows from
the last assertion of Corollary \ref{g-basic-cont-cond+c}. The
implication $\mathrm{b)}\Rightarrow \mathrm{a)}$ is proved by using
Proposition \ref{H-cont-cond}B. $\square$ \vspace{5pt}

\begin{example}\label{comp-oper-c-c+}
Let
$\mathcal{A}=\mathcal{K}^{s}_{H,h}\doteq\{\rho\in\mathfrak{S}(\mathcal{H})\,|\,\mathrm{Tr}H\rho\leq
h\}$ be the set of quantum states with "bounded mean energy" defined
by a $\mathfrak{H}$\nobreakdash-\hspace{0pt}operator $H$ with
$\mathrm{g}(H)=\inf\{\lambda>0\,|\,\mathrm{Tr}e^{-\lambda
H}<+\infty\}=0$ and positive $h$ (exceeding the minimal eigenvalue
of $H$). The von Neumann entropy is continuous on the set
$\mathcal{K}^{s}_{H,h}$ by Proposition \ref{H-cont-cond}B. The above
condition B\nobreakdash-\hspace{0pt}2 and condition $\mathrm{b)}$ in
Proposition \ref{comp-oper-c-c} show that the sufficient condition
of continuity of the function $\rho\mapsto H_{\Phi}(\rho)$ on the
set $\mathcal{K}^{s}_{H,h}$ consists in existence of a sequence
$\{h_{i}\}_{i=1}^{+\infty}$ of nonnegative numbers such that
$$
\sum_{i=1}^{+\infty}h_{i}V_{i}^{*}V_{i}\leq H\quad\textrm{and}\quad
\sum_{i=1}^{+\infty}e^{-\lambda
h_{i}}<+\infty\;\,\textrm{for}\;\,\textrm{all}\;\, \lambda>0.
$$
It is possible to show\footnote{This can be done by using the
following observation: \emph{for an arbitrary closed convex set
$\mathfrak{P}_{0}$ of probability distributions, on which the
Shannon entropy is continuous, there exists a sequence
$\{h_{i}\}_{i=1}^{+\infty}$ of nonnegative numbers such that}
$$
\sup_{\{\pi_{i}\}\in\mathfrak{P}_{0}}\sum_{i=1}^{+\infty}h_{i}\pi_{i}<+\infty\quad\textit{and}\quad
\sum_{i=1}^{+\infty}e^{-\lambda
h_{i}}<+\infty\;\,\textit{for}\;\,\textit{all}\;\, \lambda>0.
$$
This observation can be proved by using Lemma \ref{H-oper} and the
arguments from the proof of the implication
$\mathrm{(i)\Rightarrow(ii)}$ in Theorem \ref{bound-cont} based on
Dini's lemma.} that this condition is also necessary if the operator
$H$ is strictly positive and
$\mathrm{Ran}V_{i}\perp\mathrm{Ran}V_{j}$ for all sufficiently large
$i\neq j$.
\end{example}\vspace{5pt}

\subsection{The case of a converging sequence\\ of quantum
operations}

In analysis of continuity of information characteristics of a
quantum channel with respect to perturbations of this channel we
have to study continuity of the output entropy as a function of a
pair (channel, input state). Practically, the following problem
naturally arises in this analysis: for a given sequence
$\{\Phi_{n}\}$ of quantum operations in
$\,\mathfrak{F}_{\leq1}(\mathcal{H},\mathcal{H}')$ converging to an
operation $\Phi_{0}$ and for a given closed subset
$\mathcal{A}\subset\mathfrak{S}(\mathcal{H})$ to show that
\begin{equation}\label{task}
\lim_{n\rightarrow+\infty}H_{\Phi_{n}}(\rho_{n})=H_{\Phi_{0}}(\rho_{0})\quad
\forall \{\rho_{n}\}\subset\mathcal{A}\quad \textrm{such}\;
\textrm{that}\; \lim_{n\rightarrow+\infty}\rho_{n}=\rho_{0}.
\end{equation}

Summarizing the  results of Section 6 we obtain the following
observation.\vspace{5pt}

\emph{If the von Neumann entropy is continuous on the set
$\mathcal{A}$ then property (\ref{task}) holds provided one of the
following conditions is valid:}
\begin{enumerate}[1)]
\item
\emph{$\lim_{n\rightarrow+\infty}H_{\Phi_{n}}(|\varphi_{n}\rangle\langle\varphi_{n}|)=H_{\Phi_{0}}(|\varphi_{0}\rangle\langle\varphi_{0}|)$
for any sequence $\{\varphi_{n}\}$ of vectors in $\,\mathcal{H}$
converging to a vector $\varphi_{0}$;}
\item \emph{property (\ref{task}) holds for the sequence $\{\widetilde{\Phi}_{n}\}$ of quantum operations
strongly converging to the operation $\widetilde{\Phi}_{0}$ such
that $\,(\Phi_{n},\widetilde{\Phi}_{n})$ is a complementary pair for
each $n\geq0$. }\end{enumerate} \vspace{5pt}

If
$\Phi_{n}(\cdot)=\sum_{i=1}^{+\infty}V^{n}_{i}(\cdot)(V^{n}_{i})^{*}$
for each $n\geq0$, where the sequence $\{V^{n}_{i}\}_{n}$  strongly
converges to the operator $V^{0}_{i}$ for each $i$, then the above
conditions $\mathrm{1)}$ and $\mathrm{2)}$ can be replaced
respectively by the following ones:
\begin{enumerate}[1)']

\item
$$
\lim_{n\rightarrow+\infty}
H\left(\left\{\|V^{n}_{i}|\varphi_{n}\rangle\|^{2}\right\}_{i=1}^{+\infty}\right)=
H\left(\left\{\|V^{0}_{i}|\varphi_{0}\rangle\|^{2}\right\}_{i=1}^{+\infty}\right)
$$
\textit{for any sequence $\{\varphi_{n}\}$ of vectors in
$\,\mathcal{H}$ converging to a vector $\,\varphi_{0}$;}

\item
$$
\lim_{n\rightarrow+\infty}H\left(\left\{\mathrm{Tr}V^{n}_{i}\rho_{n}(V^{n}_{i})^{*}\right\}_{i=1}^{+\infty}\right)=
H\left(\left\{\mathrm{Tr}V^{0}_{i}\rho_{0}(V^{0}_{i})^{*}\right\}_{i=1}^{+\infty}\right).
$$
\end{enumerate} \vspace{5pt}
The sufficient conditions for 1') can be found in the second part of
Corollary \ref{PCE-conditions+}. The sufficient condition for 2')
looks like condition b) in Proposition \ref{comp-oper-c-c} with
$V^{n}_{i}$ instead of $V_{i}$ and $"\sup_{n\geq0}"$ added to
$"\sup_{\rho\in\mathcal{A}}"$. \vspace{5pt}

\begin{example}\label{measurement} The above condition 2') makes it possible to replace the strong*
operator topology by the strong operator topology in the assertion
in example 3 in \cite{Sh-11}, concerning quantum measurements with a
countable number of outcomes and stating that \emph{continuity of
the Shannon entropy of the outcomes probability distribution with
respect to a priori state and to a measurement implies continuity of
the von Neumann entropy of the mean posteriori state with respect to
the same variables provided a priori state varies within a set on
which the von Neumann entropy is continuous.}
\end{example}

\section{Some applications}

\subsection{The Holevo capacity of quantum channels}

Let
$\Phi:\mathfrak{T}(\mathcal{H})\mapsto\mathfrak{T}(\mathcal{H}')$ be
a quantum channel and $\mathcal{A}$ be a subset of
$\mathfrak{S}(\mathcal{H})$. The Holevo capacity of the
$\mathcal{A}$-constrained channel $\Phi$ is defined as follows
(cf.\cite{H-c-w-c, H-Sh-2})
\begin{equation}\label{ccap-1}
\bar{C}(\Phi,\mathcal{A})=\sup_{\{\pi_{i},\rho_{i}\}\in
\mathcal{P}^{\mathrm{f}}_{\mathcal{A}}(\mathfrak{S}(\mathcal{H}))}\sum_{i}\pi_{i}H(\Phi(\rho_{i})\|\Phi(\textbf{b}(\{\pi_{i},\rho_{i}\})))
\end{equation}
(the supremum is over all finite ensembles of states with the
average in $\mathcal{A}$).

\subsubsection{On existence of continuous optimal ensembles}

The well known fact concerning the Holevo capacity of a finite
dimensional quantum channel $\Phi$ constrained by a closed subset
$\mathcal{A}$ consists in existence of an optimal ensemble at which
the supremum in (\ref{ccap-1}) is achieved \cite{Schum-West}. Since
\begin{equation}\label{ccap-2}
\bar{C}(\Phi,\mathcal{A})=\sup_{\mu\in
\mathcal{P}_{\mathcal{A}}(\mathfrak{S}(\mathcal{H}))}\int_{\mathfrak{S}(\mathcal{H})}H(\Phi
(\rho )\Vert \Phi (\textbf{b}(\mu)))\mu(d\rho )
\end{equation}
(the supremum is over all probability measures with the barycenter
in $\mathcal{A}$) the notion of an optimal ensemble is naturally
generalized to the infinite dimensional case leading to the notion
of an optimal measure (generalized  or continuous optimal ensemble)
at which the supremum in (\ref{ccap-2}) is achieved \cite{H-Sh-2}.
In contrast to the finite dimensional case we can not claim
existence of an optimal measure for an arbitrary quantum channel
constrained by closed or even compact subset of states (see the
example in \cite{H-Sh-2}). By the theorem in \cite{H-Sh-2},
containing a sufficient condition for existence of an optimal
measure, we have the following result.\vspace{5pt}
\begin{property}\label{application-2}
\textit{Let
$\,\Phi:\mathfrak{T}(\mathcal{H})\rightarrow\mathfrak{T}(\mathcal{H}')$
be a quantum channel and let $\mathcal{A}$ be a compact subset of
$\,\mathfrak{S}(\mathcal{H})$. If one of the conditions of
continuity of the function $\mathcal{A}\ni\rho\mapsto
H_{\Phi}(\rho)$ presented in Section 7.1 holds then there exists an
optimal measure for the $\mathcal{A}$-constrained channel $\,\Phi$
supported by pure states.}
\end{property}
\vspace{5pt}

This proposition implies existence of an optimal measure for an
arbitrary quantum channel with finite dimensional environment (with
finite Kraus representation) constrained by a compact subset of
states on which the entropy is continuous.

\subsubsection{On continuity of the Holevo capacity\\ as a function of a
channel}

In the finite dimensional case the Holevo capacity
$\bar{C}(\Phi,\mathcal{A})$ is a continuous function of $\Phi$ on
the set of all quantum channels
$\mathfrak{F}_{=1}(\mathcal{H},\mathcal{H}')$, but it is not
continuous in infinite dimensions even with respect to the norm of
complete boundedness \cite{L&S}. By proposition 7 in \cite{Sh-H},
containing a  sufficient condition for continuity of the function
$\,\Phi\mapsto\bar{C}(\Phi,\mathcal{A})\,$ on subsets of
$\,\mathfrak{F}_{=1}(\mathcal{H},\mathcal{H}')\,$ with respect to
the topology of strong convergence, we have the following result.
\vspace{5pt}

\begin{property}\label{application-1}
\textit{Let $\,\{\Phi_{n}\}$ be a sequence of quantum channels
strongly converging to a quantum channel $\,\Phi_{0}$ and let
$\mathcal{A}$ be a  compact subset of $\,\mathfrak{S}(\mathcal{H})$
on which the von Neumann entropy is continuous. If one of the
conditions of validity of (\ref{task}) presented in Section 7.2
holds then
$$
\lim_{n\rightarrow+\infty}\bar{C}(\Phi_{n},\mathcal{A})=\bar{C}(\Phi_{0},\mathcal{A}).
$$}
\end{property}

\begin{corollary}\label{application-1+}
\textit{For arbitrary compact subset $\mathcal{A}$ of
$\,\mathfrak{S}(\mathcal{H})$, on which the von Neumann entropy is
continuous, the function $\Phi\mapsto\bar{C}(\Phi,\mathcal{A})$ is
continuous on the set
$\,\mathfrak{F}^{k}_{=1}(\mathcal{H},\mathcal{H}')$ for each
$k$.\footnote{$\mathfrak{F}^{k}_{=1}(\mathcal{H},\mathcal{H}')$ is
the set of all quantum channels from $\mathfrak{T}(\mathcal{H})$ to
$\mathfrak{T}(\mathcal{H}')$ having the Kraus representation
consisting of $\,\leq k$ summands.}}
\end{corollary}\vspace{5pt}

Note that the set of channels having  finite Kraus representation is
dense in the set $\mathfrak{F}_{=1}(\mathcal{H},\mathcal{H}')$ of
all channels in the topology of strong convergence.

\subsection{On continuity of the Entanglement of Formation}

The notion of the Entanglement of Formation as a quantitative
characteristic of entanglement of a state in a composite quantum
system is introduced in \cite{B&Co} in the finite dimensional case.
The possible infinite dimensional generalizations of this notion are
based respectively on the $\sigma$-convex roof and the $\mu$-convex
roof constructions \cite{ESP,Sh-9}. Comparison of these
constructions, in particular, the sufficient conditions for their
coincidence on subsets of states of composite system are presented
in \cite{Sh-9}, where the arguments showing preferability of the
$\mu$-convex roof construction are also considered.\footnote{The
question of coincidence of the $\mu$-convex roof and the
$\sigma$-convex roof constructions of the EoF on the whole state
space is open (as far as I know). In \cite{Sh-9} it is shown that
this question can not be solved by using only such general
properties of the entropy as concavity and lower semicontinuity
since the $\sigma$-convex roof construction applied to a concave
lower semicontinuous and even bounded function instead of the
entropy may be different from the corresponding $\mu$-convex roof
construction (and may not satisfy the basic condition of
entanglement monotones).} So, in what follows we will use the
generalization of the EoF based on this construction, t.i.
$$
E_{F}(\omega)=(H\circ\Theta)_{*}^{\mu}(\omega)\doteq\inf
\int_{\mathrm{extr}\mathfrak{S}(\mathcal{H}\otimes\mathcal{K})}H(\Theta(\varpi))\mu(d\varpi),
\quad\omega\in\mathfrak{S}(\mathcal{H}\otimes\mathcal{K}),
$$
where $\Theta(\cdot)=\mathrm{Tr}_{\mathcal{K}}(\cdot)$  and the
infimum is over all probability measures $\mu$ on the set
$\mathrm{extr}\mathfrak{S}(\mathcal{H}\otimes\mathcal{K})$ with the
barycenter $\omega$.\vspace{5pt}

By proposition 8 in \cite{Sh-9} to show continuity of the function
$\omega\mapsto E_{F}(\omega)$ on a particular subset of
$\mathfrak{S}(\mathcal{H}\otimes\mathcal{K})$ it is sufficient to
show continuity of one of the functions $\omega\mapsto
H(\mathrm{Tr}_{\mathcal{K}}\omega)$ and $\omega\mapsto
H(\mathrm{Tr}_{\mathcal{H}}\omega)$ on this subset. Thus by using
the results of the previous sections one can obtain continuity
conditions for the function $\omega\mapsto E_{F}(\omega)$.
\vspace{5pt}

\begin{property}\label{EoF-2}
\textit{Let $\,\{\omega_{n}\}$ be a sequence of states in
$\,\mathfrak{S}(\mathcal{H}\otimes\mathcal{K})$, converging to a
state $\omega_{0}$, such that
$\,\lim_{n\rightarrow+\infty}H(\rho_{n})=H(\rho_{0})<+\infty$, where
$\rho_{n}=\mathrm{Tr}_{\mathcal{K}}\omega_{n}$ for $n=0,1,2..$. Let
$\,\{\Phi_{n}\}$ and $\,\{\Psi_{n}\}$ be sequences of operations in
$\,\mathfrak{F}_{\leq 1}(\mathcal{H})$ and in $\,\mathfrak{F}_{\leq
1}(\mathcal{K})$ strongly converging to operations $\,\Phi_{0}$ and
$\,\Psi_{0}$ correspondingly. If one of the conditions of validity
of (\ref{task}) for the sequences $\{\rho_{n}\}$ and
$\,\{\Phi_{n}\}$ presented in Section 7.2 holds and
$\,\mathrm{Tr}\,\Phi_{0}\otimes\Psi_{0}(\omega_{0})>0$ then}
$$
\lim_{n\rightarrow+\infty}E_{F}\left(\frac{\Phi_{n}\otimes\Psi_{n}(\omega_{n})}{\mathrm{Tr}\,\Phi_{n}\otimes\Psi_{n}(\omega_{n})}\right)
=E_{F}\left(\frac{\Phi_{0}\otimes\Psi_{0}(\omega_{0})}{\mathrm{Tr}\,\Phi_{0}\otimes\Psi_{0}(\omega_{0})}\right).
$$
\end{property}
\vspace{5pt}

\textbf{Proof.} By the remark before the proposition it is
sufficient to show that
$$
\lim_{n\rightarrow+\infty}H(\mathrm{Tr}_{\mathcal{K}}\Phi_{n}\otimes\Psi_{n}(\omega_{n}))=
H(\mathrm{Tr}_{\mathcal{K}}\Phi_{0}\otimes\Psi_{0}(\omega_{0}))<+\infty.
$$
Since by the condition we have
$\;\lim_{n\rightarrow+\infty}H(\Phi_{n}(\rho_{n}))=H(\Phi_{0}(\rho_{0}))<+\infty\;$
and
$\mathrm{Tr}_{\mathcal{K}}\Phi_{n}\otimes\Psi_{n}(\omega_{n})\leq\Phi_{n}(\rho_{n})$
for $n=0,1,2,...$, the above relation follows from corollary 4 in
\cite{Sh-11}. $\square$\vspace{5pt}

The assertion of this proposition is valid in the following
cases:\begin{itemize}
    \item $\Phi_{n}=\Phi_{0}$ is a PCE\nobreakdash-\hspace{0pt}operation (see Section 4);
    \item $\{\Phi_{n}\}\subset\mathfrak{F}^{k}_{\leq
1}(\mathcal{H})$ (the set of quantum operations having the Kraus
representation consisting of $\,\leq k\,$ summands).
\end{itemize}

Proposition \ref{EoF-2} can be used in analysis of continuity of the
EoF under local operations.

\section{Appendix}

\subsection{Continuity of the entropy on subsets of $\mathfrak{T}_{+}(\mathcal{H}\otimes\mathcal{K})$}

Proposition 10 in \cite{Sh-4} can be generalized to subsets of
$\mathfrak{T}_{+}(\mathcal{H}\otimes\mathcal{K})$ as
follows.\vspace{5pt}

\begin{property}\label{cont-cond-a-1}
\textit{Let $\mathcal{C}$ be a subset of
$\,\mathfrak{T}_{+}(\mathcal{H}\otimes\mathcal{K})$. Continuity of
the quantum entropy on the sets
$\mathrm{Tr}_{\mathcal{K}}\mathcal{C}\subset\mathfrak{T}_{+}(\mathcal{H})$
and
$\mathrm{Tr}_{\mathcal{H}}\mathcal{C}\subset\mathfrak{T}_{+}(\mathcal{K})$
implies continuity of the quantum entropy on the set
$\,\mathcal{C}$.}
\end{property}\vspace{5pt}
\textbf{Proof.} Let $\{C_{n}\}\subseteq\mathcal{C}$ be a sequence
converging to an operator $C_{0}\in\mathcal{C}$. If $C_{0}\neq0$
then by the condition we have
$$
\lim_{n\rightarrow+\infty}H\left(\frac{C_{n}^{\mathcal{H}}}{\mathrm{Tr}C_{n}}\right)=
H\left(\frac{C_{0}^{\mathcal{H}}}{\mathrm{Tr}C_{0}}\right)\quad\textrm{and}\quad
\lim_{n\rightarrow+\infty}H\left(\frac{C_{n}^{\mathcal{K}}}{\mathrm{Tr}C_{n}}\right)=
H\left(\frac{C_{0}^{\mathcal{K}}}{\mathrm{Tr}C_{0}}\right)
$$
where $C_{n}^{\mathcal{H}}=\mathrm{Tr}_{\mathcal{K}}C_{n}$ and
$C_{n}^{\mathcal{K}}=\mathrm{Tr}_{\mathcal{H}}C_{n}$ for all $n$.
Proposition 10 in \cite{Sh-4} implies
$$
\lim_{n\rightarrow+\infty}H\left(\frac{C_{n}}{\mathrm{Tr}C_{n}}\right)=
H\left(\frac{C_{0}}{\mathrm{Tr}C_{0}}\right)\quad\textrm{and
 hence}\quad \lim_{n\rightarrow+\infty}H\left(C_{n}\right)=
H\left(C_{0}\right).
$$
If $C_{0}=0$ then convergence to zero of the sequence
$\{H\left(C_{n}\right)\}$ follows from convergence to zero of the
sequences $\{H\left(C_{n}^{\mathcal{H}}\right)\}$ and
$\{H\left(C_{n}^{\mathcal{K}}\right)\}$ by means of the inequality
$H\left(C_{n}\right)\leq
H\left(C_{n}^{\mathcal{H}}\right)+H\left(C_{n}^{\mathcal{K}}\right)$,$\;\,n\in\mathbb{N}$.
$\square$

\subsection{Auxiliary results}

\begin{lemma}\label{nv-2}
\emph{Let $\,\{\rho_{n}\}$ be a sequence of states converging to a
state $\rho_{0}$ such that
$\,\lim_{n\rightarrow+\infty}H(\rho_{n})=H(\rho_{0})$. For given
natural $k$ let $P^{k}_{n}$ be a $k$\nobreakdash-\hspace{0pt}rank
spectral projector of the state $\rho_{n}$ corresponding to its $k$
maximal eigenvalues and let $A^{k}_{n}=P^{k}_{n}\rho_{n}$ for all
$\,n$.\footnote{The projector $P^{k}_{n}$ is uniquely defined if the
state $\rho_{n}$ has no multiple eigenvalues. In any case all
variants of $P^{k}_{n}$ lead to isomorphic operators $A^{k}_{n}$ and
we assume that one of them is chosen.} For arbitrary $\varepsilon>0$
there exist a natural $k_{\varepsilon}$ and a subsequence
$\{A^{k_{\varepsilon}}_{n_{t}}\}$ converging to the operator
$A^{k_{\varepsilon}}_{0}=\bar{P}^{k_{\varepsilon}}_{0}\rho_{0}$,
where $\bar{P}^{k_{\varepsilon}}_{0}$ is a particular
$k_{\varepsilon}$\nobreakdash-\hspace{0pt}rank spectral projector of
the state $\rho_{0}$ corresponding to its $k_{\varepsilon}$ maximal
eigenvalues, such that
$$
\mathrm{Tr}B^{k_{\varepsilon}}_{n_{t}}<\varepsilon\quad
\textit{and}\quad H(B^{k_{\varepsilon}}_{n_{t}})<\varepsilon,
$$
where
$B^{k_{\varepsilon}}_{n_{t}}=\rho_{n_{t}}-A^{k_{\varepsilon}}_{n_{t}}$
is a positive operator, for all sufficiently large $t$.}
\end{lemma}\vspace{5pt}
\textbf{Proof.} Despite possible multiple meaning of the operator
$P^{k}_{0}\rho_{0}$ the values $\mathrm{Tr}P^{k}_{0}\rho_{0}$ and
$H(P^{k}_{0}\rho_{0})$ are uniquely defined by the state $\rho_{0}$.
By lemma 4 in \cite{L-2} the sequence $\{H(P^{k}_{0}\rho_{0})\}_{k}$
is nondecreasing and tends to $H(\rho_{0})$ as
$k\rightarrow+\infty$. Let $k_{\varepsilon}$ be such that
$\mathrm{Tr}P^{k_{\varepsilon}}_{0}\rho_{0}>1-\varepsilon/2$ and
$H(\rho_{0})-H(P^{k_{\varepsilon}}_{0}\rho_{0})<\varepsilon/3$.
Since $A^{k_{\varepsilon}}_{n}\leq\rho_{n}$ for all $n$, the
compactness criterion for subsets of $\mathfrak{T}_{+}(\mathcal{H})$
(see the Appendix in \cite{Sh-11}) shows relative compactness of the
sequence $\{A^{k_{\varepsilon}}_{n}\}$ and hence existence of a
subsequence $\{A^{k_{\varepsilon}}_{n_{t}}\}$ converging to an
operator $A^{k_{\varepsilon}}_{0}$. By using coincidence of
$\mathrm{Tr}A^{k_{\varepsilon}}_{n}$ with
$\sup_{P}\mathrm{Tr}P\rho_{n}$, where $P$ runs over the set of all
$k_{\varepsilon}$\nobreakdash-\hspace{0pt}rank projectors, it is
easy to show that
$A^{k_{\varepsilon}}_{0}=\bar{P}^{k_{\varepsilon}}_{0}\rho_{0}$,
where $\bar{P}^{k_{\varepsilon}}_{0}$ is a particular
$k_{\varepsilon}$\nobreakdash-\hspace{0pt}rank spectral projector of
the state $\rho_{0}$ corresponding to its $k_{\varepsilon}$ maximal
eigenvalues. \vspace{5pt}

Since
$\,\lim_{t\rightarrow+\infty}\mathrm{Tr}A^{k_{\varepsilon}}_{n_{t}}=\mathrm{Tr}A^{k_{\varepsilon}}_{0}\,$
and
$\,\lim_{t\rightarrow+\infty}H(A^{k_{\varepsilon}}_{n_{t}})=H(A^{k_{\varepsilon}}_{0})\,$
we have
$$
\begin{array}{c}
\mathrm{Tr}B^{k_{\varepsilon}}_{n_{t}}=
1-\mathrm{Tr}A^{k_{\varepsilon}}_{n_{t}}\leq
|1-\mathrm{Tr}A^{k_{\varepsilon}}_{0}|+
|\mathrm{Tr}A^{k_{\varepsilon}}_{n_{t}}-\mathrm{Tr}A^{k_{\varepsilon}}_{0}|<\varepsilon/2+\varepsilon/2=\varepsilon
\end{array}
$$
and
$$
\begin{array}{c}
H(B^{k_{\varepsilon}}_{n_{t}})\leq H(\rho_{n_{t}})-
H(A^{k_{\varepsilon}}_{n_{t}})\leq
|H(\rho_{n_{t}})-H(\rho_{0})|\\\\+|H(\rho_{0})-H(A^{k_{\varepsilon}}_{0})|+
|H(A^{k_{\varepsilon}}_{n_{t}})-H(A^{k_{\varepsilon}}_{0})|<\varepsilon/3+\varepsilon/3+\varepsilon/3=\varepsilon
\end{array}
$$
for all sufficiently large $t$, where the second inequality is
obtained by using inequality (\ref{H-fun-ineq}).
$\square$\vspace{5pt}

\begin{lemma}\label{simple+}
\textit{Let $\{\pi_{i}\}_{i=1}^{+\infty}$ be a sequence of positive
numbers. Then
$$
\sup_{\{x_{i}\}\in\mathfrak{P}_{+\infty}}H\left(\left\{\pi_{i}x_{i}\right\}_{i=1}^{+\infty}\right)=\lambda^{*},
$$
where $\lambda^{*}$ is either the unique finite solution of the
equation $\sum_{i=1}^{+\infty}e^{-\lambda/\pi_{i}}=1$ if it exists
or equal to
$\,\mathrm{g}(\{\pi^{-1}_{i}\})=\inf\{\lambda>0\,|\,\sum_{i=1}^{+\infty}e^{-\lambda/\pi_{i}}<+\infty\}$
otherwise.}\footnote{It is assumed that $\inf\emptyset=+\infty$. The
equation $\sum_{i=1}^{+\infty}e^{-\lambda/\pi_{i}}=1$ has no
solution if and only if either
$\mathrm{g}(\{\pi^{-1}_{i}\})=+\infty$ or
$\sum_{i=1}^{+\infty}e^{-\mathrm{g}(\{\pi^{-1}_{i}\})/\pi_{i}}<1$.}
\end{lemma}\vspace{5pt}

\textbf{Proof.} By using the Lagrange method it is easy to show that
the function $\mathfrak{P}_{n}\ni\{x_{i}\}_{i=1}^{n}\mapsto
H\left(\left\{\pi_{i}x_{i}\right\}_{i=1}^{n}\right)$ attains its
maximum  at the vector
$\left\{x^{*}_{i}=c\pi^{-1}_{i}e^{-\lambda^{*}_{n}/\pi_{i}}\right\}$,
where $\lambda^{*}_{n}$ is the  solution of the equation
$\sum_{i=1}^{n}e^{-\lambda/\pi_{i}}=1$ and
$c=\left[\sum_{i=1}^{n}\pi^{-1}_{i}e^{-\lambda^{*}_{n}/\pi_{i}}\right]^{-1}$.
Hence \begin{equation}\label{double-ineq}
\max_{\{x_{i}\}\in\mathfrak{P}_{n}}H\left(\left\{\pi_{i}x_{i}\right\}_{i=1}^{n}\right)=\lambda^{*}_{n}.
\end{equation}

The assertion of the lemma can be derived from (\ref{double-ineq})
by noting that the sequence $\{\lambda^{*}_{n}\}$ tends to
$\lambda^{*}$ as $n\rightarrow+\infty$ and by using lower
semicontinuity of the classical entropy.  $\square$\vspace{15pt}

I am grateful to A.S.Holevo and the participants of his seminar for
the useful discussion. I am also grateful to the organizers of the
workshop \emph{Thematic Program on Mathematics in Quantum
Information} at the Fields Institute, where the work on this paper
was initiated. This work is partially supported by the program
"Mathematical control theory" of Russian Academy of Sciences, by the
federal target program "Scientific and pedagogical staff of
innovative Russia" (program 1.2.1, contract P 938), by the
analytical departmental target program "Development of scientific
potential of the higher school 2009-2010" (project 2.1.1/500) and by
RFBR grant 09-01-00424-a.

\end{document}